\documentclass[fleqn,usenatbib]{mnras}
\usepackage{siunitx}
\usepackage{physics}
\usepackage{graphicx}
\usepackage{multirow}
\usepackage{pifont}
\usepackage{appendix}
\usepackage{caption}
\usepackage{subcaption}
\usepackage{supertabular}
\usepackage{rotating}

\newcommand{\cmark}{\ding{51}}

\DeclareSIUnit\jansky{Jy}
\DeclareSIUnit\parsec{pc}
\DeclareSIUnit\year{yr}
\DeclareSIUnit\erg{erg}

\title[CHIME FRB supervised machine learning]{Machine learning classification of CHIME fast radio bursts -- I. Supervised Methods}
\author[J-W Luo et. al]{Jia-Wei Luo$^{1,2}$\thanks{luoj7@unlv.nevada.edu}, Jia-Ming Zhu-Ge$^{3}$ and Bing Zhang$^{1,2}$
\\
$^{1}$Nevada Center for Astrophysics, University of Nevada, Las Vegas, NV 89154, USA\\
$^{2}$Department of Physics and Astronomy, University of Nevada, Las Vegas, NV 89154, USA\\
$^{3}$Department of Astronomy, University of Science and Technology of China, Hefei 230026, China\\
}

\date{Accepted XXX. Received YYY; in original form ZZZ}
\pubyear{2022}
\begin{document}
\label{firstpage}
\pagerange{\pageref{firstpage}--\pageref{lastpage}}
\maketitle

\begin{abstract}
Observationally, the mysterious fast radio bursts (FRBs) are classified as repeating ones and apparently non-repeating ones. While repeating FRBs cannot be classified into the non-repeating group, it is unknown whether the apparently non-repeating FRBs are actually repeating FRBs whose repetitions are yet to be discovered, or whether they belong to another physically distinct type from the repeating ones. In a series of two papers, we attempt to disentangle this mystery with machine learning methods. In this first paper, we focus on an array of supervised machine learning methods. We train the machine learning algorithms with a fraction of the observed FRBs in the first CHIME/FRB catalog, telling them which ones are apparently non-repeating and which ones are repeating. We then let the trained models predict the repetitiveness of the rest of the FRB data with the observed parameters, and we compare the predictions with the observed repetitiveness. We find that the models can predict most FRBs correctly, hinting towards distinct mechanisms behind repeating and non-repeating FRBs. We also find that the two most important distinguishing factors between non-repeating and repeating FRBs are brightness temperature and rest-frame frequency bandwidth. By applying the trained models back to the entire first CHIME catalog, we further identify some potentially repeating FRBs currently reported as non-repeating. We recommend a list of these bursts as targets for future observing campaigns to search for repeated bursts in a combination with the results presented in Paper II using unsupervised machine learning methods.
\end{abstract}


\section{Introduction}
\label{sec:introduction}

Fast radio bursts (FRBs) are a type of millisecond luminous cosmic bursts in the radio wavelength  \citep{lorimer2007BrightMillisecondRadio,thornton2013PopulationFastRadio,petroff2022FastRadioBursts,cordes2019FastRadioBursts} whose origins have yet been brought to light \citep{platts2019LivingTheoryCatalogue,zhang2020PhysicalMechanismsFast,xiao2022FastRadioBursts}. Many models on the origin of FRBs have been proposed, albeit not a single one of them is able to explain all FRB phenomena \citep[for reviews, see e.g.][]{katz2016FastRadioBursts,platts2019LivingTheoryCatalogue,zhang2020PhysicalMechanismsFast,xiao2022FastRadioBursts}. 

While most FRBs are observed as one-off events, a small fraction, the first of which is FRB 20121102 \citep{spitler2016RepeatingFastRadio}, show repeated bursts with a range of bursting rates. These repeating FRBs eliminated the possibility of cataclysmic event models being applied to all FRBs. However, an argument can still be made that repeating and non-repeating FRBs originate from different progenitors, and that only repeating FRBs rise from repeatable mechanisms \citep[for discussions, see e.g.][]{katz2022SourcesFastRadio}.

Prior to testing which physical model can explain the proprieties of the two potential types of FRBs, the existence of such dichotomy must first be proved \citep{palaniswamy2018AreThereMultiple,ravi2019PrevalenceRepeatingFast,caleb2019AreAllFast,lu2020ImplicationsCanadianHydrogen,ai2021TrueFractionsRepeating,guo2022PossibleSubclassificationFast,zhong2022CanSinglePopulation}. There are some studies on the differences between repeating and non-repeating FRBs, pointing out some difference in parameter distributions of repeating and non-repeating FRBs such as burst width, bandwidth, spectral shape, energy and brightness temperature \citep{andersen2019CHIMEFRBDiscovery,fonseca2020NineNewRepeating,li2021BimodalBurstEnergy,aggarwal2021ObservationalEffectsBanded,chime/frbcollaboration2021FirstCHIMEFRB,pleunis2021FastRadioBurst,li2021LongShortFast,xiao2022NewInsightsCriterion,zhang2022StatisticalSimilarityRepeating,cui2022LuminosityDistributionFast}. However, these comparisons usually only consider differences within a single set of parameters, and cannot uncover discrepancies in combination of parameters in a higher dimensional parameter space.

A way to mitigate this limitation is through machine learning. Machine learning algorithms can analyze a large amount of data simultaneously and are capable of highlighting patterns in very high-dimensional data with minimal human intervention \citep[for reviews on the application of machine learning in astronomy, see e.g.][]{ball2012DATAMININGMACHINE,baron2019MachineLearningAstronomy,ivezic2014StatisticsDataMining}.

There are two main classification approaches with machine learning: supervised and unsupervised. Supervised machine learning algorithms are used to learn a mapping from a set of input features to a set of labels, with the latter being inputted by human experts. After training, supervised machine learning algorithms can automatically label new input data without human input. Unsupervised machine learning algorithms, on the other hand, only take features as input and do not take labels. The unsupervised machine learning algorithm only knows the appearance of the data, but does not know which class each data point belongs to. It attempts to group input data points that looked similar into different clusters, without knowing or explaining the physical relationship of the clusters.

While machine learning methods are already widely used in aiding discovery of FRBs \citep[see e.g.][]{wagstaff2016MachineLearningClassifier,zhang2018FastRadioBursta,connor2018ApplyingDeepLearning,thechime/frbcollaboration2019ObservationsFastRadio,wu2019FeatureMatchingConditional,farah2019FiveNewRealtime,adamek2020SinglepulseDetectionAlgorithms,agarwal2020FETCHDeeplearningBased,yang202181NewCandidate}, few 
studies have employed machine learning methods to the classification of previously discovered FRBs, especially with supervised machine learning methods. \citet{chen2021UncloakingHiddenRepeating} used unsupervised machine learning methods to find apparent non-repeating FRBs that share similar observed parameters with the apparent repeating ones, and identified some repeater candidates. \citet{chaikova2022ModelindependentClassificationEvents}, on the other hand, found two clusters of FRBs with distinct properties with similar unsupervised machine learning methods. 

Supervised machine learning differs from unsupervised methods used by above-mentioned studies in that the supervised algorithms are able to make deterministic predictions of individual FRBs instead of clusters of FRBs, and can thus be more useful in identifying repeater candidates.

In this paper, we utilize a wide array of supervised machine learning algorithms implemented in \texttt{scikit-learn} \citep{pedregosa2011ScikitlearnMachineLearning} and other open-source libraries and apply them to the classification of repeating and non-repeating FRBs. In Section \ref{sec:data}, we describe the input data and features we utilize and construct. In Section \ref{sec:methods}, we introduce the data pre-processing procedure and model evaluation methods. In Section \ref{sec:results}, we report the machine learning models and the outputs from them. We also compare the prediction accuracy of different models, as well as the importance of different observational parameters to the binary classification of repeating and non-repeating FRBs. In Section \ref{sec:uncover_repeaters}, we apply the trained models back to the entire dataset and attempt to uncover hidden repeating FRBs. In Section \ref{sec:conclusions}, we draw our conclusions. In a companion paper \citep{zhu-ge2022MachineLearningClassification}, we conduct a set of unsupervised machine learning methods to classify the same data set. The consistency between the two methods is checked and the overlapping repeating FRB candidates are reported.

\section{Data}
\label{sec:data}

We use the first CHIME/FRB catalog \citep{chime/frbcollaboration2021FirstCHIMEFRB}, which has a total of 536 FRBs observed between July 25, 2018 and July 1, 2019. Among them are 474 non-repeating bursts, and 62 bursts from 18 repeating FRB sources. We treat each sub-burst as an independent burst, and we exclude 6 bursts without flux measurements (FRB20190307A, FRB20190307B, FRB20190329B, FRB20190329C, FRB20190531A, FRB20190531B, all non-repeating). This leaves us with 594 individual bursts including sub-bursts, consisting of 500 bursts from the apparently non-repeating category and 94 bursts from the repeating category.

The input features to the machine learning algorithms in this study can be classified into two types: primary features that are drawn directly from the CHIME catalog, and secondary features that are calculated from primary features.

\begin{figure*}
    \centering
	\includegraphics[width=0.99\textwidth]{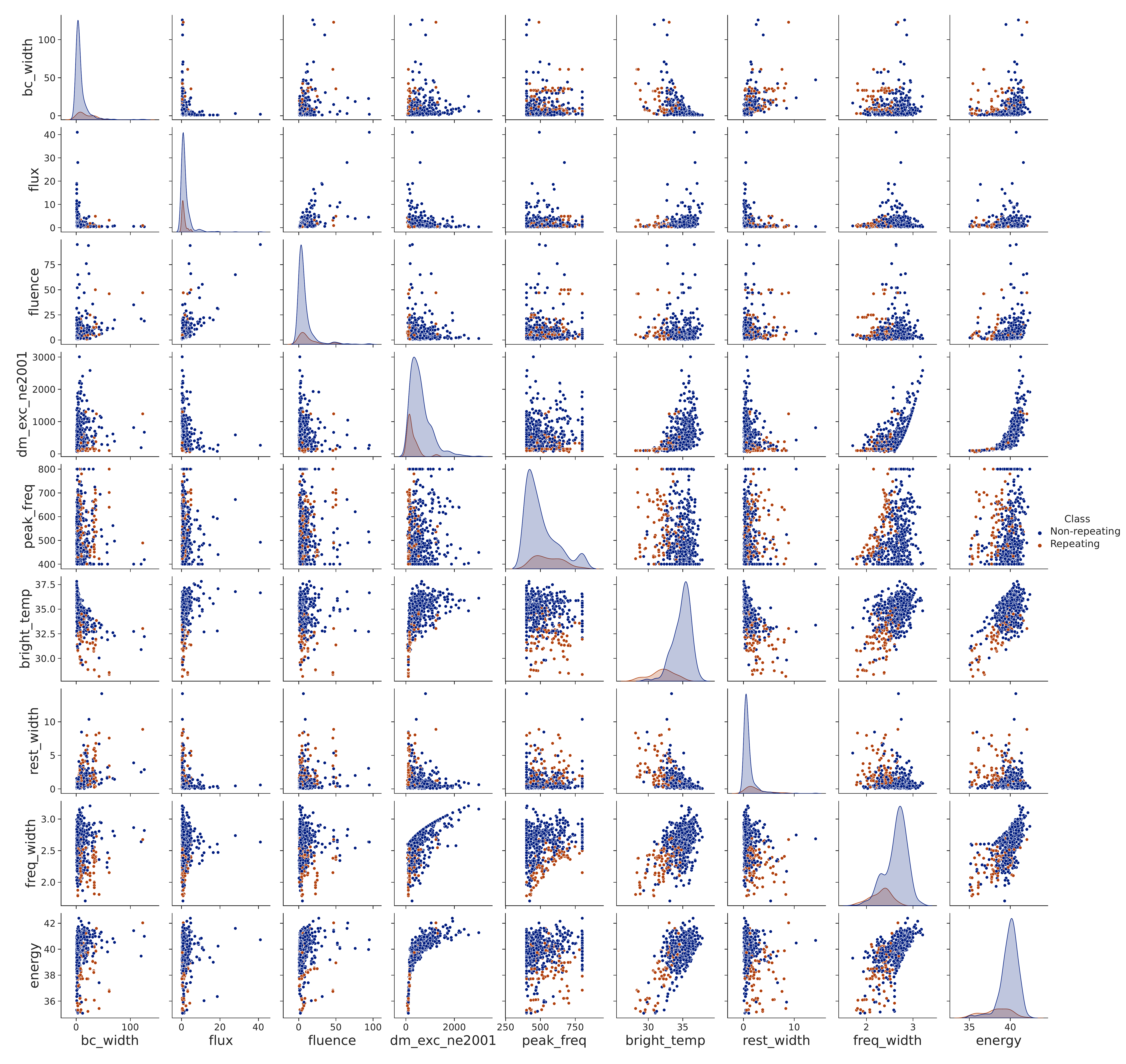}
	\caption{Corner plot of input feature distributions. The apparently non-repeating and repeating samples are denoted in blue and orange colors, respectively.}
	\label{fig:features}
\end{figure*}

The distributions of the input features are shown in Fig. \ref{fig:features}, with the details discussed below:

\subsection{Primary features}
\label{subsec:primary_features}

\begin{itemize}
    \item Box-car width ($\si{\s}$)
    
    This width represents a rough measure of the duration of the entire burst combining all sub-bursts \citep[][the same below]{chime/frbcollaboration2021FirstCHIMEFRB}. The CHIME catalog also provides a more sophisticated \texttt{fitburst} width, which we utilize to calculate rest widths later in this section.
    
    \item Flux ($\si{\jansky}$)
    
    The flux reported by the CHIME catalog is the peak flux averaged across the frequency band. For bursts with sub-bursts, the CHIME catalog provides one single flux and we adopt the same value for all of the sub-bursts.
    
    \item Fluence ($\si{\jansky\milli\s}$)
    
    The Fluence reported by the CHIME catalog is the integral of the flux time series across the burst extent and averaged across the frequency band. This value is also the same for sub-bursts in a burst as provided by the CHIME catalog.
    
    \item Excess dispersion measure ($\si{\parsec\per\cubic\centi\m}$)
    
    Excess dispersion measure (DM) is the DM of the FRB excluding the galaxy disk component. We use the NE2001 electron-density model \citep{cordes2003NE2001NewModel} values provided by the CHIME collaboration. This value is the same for sub-bursts.
    
    \item Peak frequency ($\si{\mega\hertz}$)
    
    Peak frequency is the frequency corresponding to the highest flux density. This value is different for different sub-bursts.
    
\end{itemize}

The CHIME catalog provides scattering time as a representation of scattering of an FRB, but due to the degeneracy with the intrinsic shape of the burst, this parameter is given as an upper limit if the confidence level is not high enough from the CHIME FRB fitting routine \texttt{fitbutst} \citep{chime/frbcollaboration2021FirstCHIMEFRB,chawla2022ModelingFastRadio}. It is not straightforward to incorporate upper limits together with other values for machine learning algorithms, so we do not use scattering time in this study. 

The CHIME catalog also provides spectral information of FRBs as spectral index and spectral running fitted with a powerlaw with running model \citep{chime/frbcollaboration2021FirstCHIMEFRB,chawla2022ModelingFastRadio}. We found this universal model difficult to apply as some FRBs are fitted with spectral indices of $\sim 50$. Thus we do not include these two features in our model. The spectral properties of FRBs are represented by peak frequency and frequency bandwidth.

\subsection{Secondary features}
\label{subsec:secondary_features}

\begin{itemize}
    \item Redshift
    
    In the CHIME 1st catalog, only two sources have measured redshifts through localization. In these two cases, we adopt their true redshifts: 
    FRB20121102A at $z=0.19273$ and FRB20180916B at $z=0.0337$ \citep{tendulkar2017HostGalaxyRedshift,marcote2020RepeatingFastRadio}.

    For the majority of bursts without directly measured redshifts, we use the observed DMs to estimate $z$ following the standard procedure. In general, the observed DM of an FRB can be broken down into four components \citep[e. g.][]{thornton2013PopulationFastRadio,deng2014COSMOLOGICALIMPLICATIONSFAST,prochaska2019ProbingGalacticHalos}:
    \begin{equation}
        {\rm DM=DM_{MW}+DM_{Halo}+DM_{IGM}}+\frac{\rm DM_{Host}}{1+z},
    \end{equation}
    where $\rm DM_{MW}$ is the contribution from the Milky way disk, $\rm DM_{Halo}$ is the contribution from the Milky way halo, $\rm DM_{IGM}$ is the contribution from inter-galactic medium (IGM), and $\rm DM_{Host}$ is the contribution from the FRB host galaxy. The observed host galaxy DM is decreased by a factor of $(1+z)$ due to cosmological effects.
    
    Among these four components, $\rm DM_{IGM}$ is directly tied to the redshift of the FRB through (\citealt{deng2014COSMOLOGICALIMPLICATIONSFAST,gao2014FASTRADIOBURST,zhou2014FastRadioBursts,macquart2020CensusBaryonsUniverse,li2020CosmologyinsensitiveEstimateIGM,cui2022LuminosityDistributionFast}, see also \cite{ioka2003CosmicDispersionMeasure,inoue2004ProbingCosmicReionization}):
    \begin{equation}
        {\rm DM_{IGM}}(z) = \frac{3cH_0\Omega_b f_{\rm IGM}}{8\upi G m_p}\int_0^z \frac{\chi(z)(1+z)}{\sqrt{\Omega_m(1+z)^3+\Omega_{\Lambda}}}\,\dd z
    \end{equation}
    Where $c$ is the speed of light, $G$ is the gravitational constant, $\chi(z)$ is a factor denoting ionized electrons to baryons in the IGM, which $\sim 7/8$ if both H and He are fully ionized, $m_p$ is the mass of protons, and $f_{\rm IGM}$ is the fraction of baryons in the IGM. We assume $\chi$ to be a constant of $7/8$, and $f_{IGM}\sim0.83$ \citep{fukugita1998CosmicBaryonBudget,li2020CosmologyinsensitiveEstimateIGM} and adopt the cosmology parameters measured by Planck \citep{aghanim2020Planck2018Results}: $H_0=\SI{67.4}{\kilo\meter\per\second\per\mega\parsec}$, $\Omega_m=0.315$, $\Omega_b=0.0493$ and $\Omega_{\Lambda}=0.685$.
    
    To obtain $\rm DM_{IGM}$, one must first deduct the other components from the observed DM. For $\rm DM_{MW}$ we use the NE2001 electron-density model \citep{cordes2003NE2001NewModel}. We assume $\rm DM_{Halo}$ to be a constant of $\SI{30}{\parsec\per\cubic\centi\m}$, and $\rm DM_{Host} = \SI{70}{\parsec\per\cubic\centi\m}$ following other studies \citep{xu2015ExtragalacticDispersionMeasures,dolag2015ConstraintsDistributionEnergetics,shannon2018DispersionBrightnessRelation,li2020CosmologyinsensitiveEstimateIGM,yamasaki2020GalacticHaloContribution,hashimoto2020NoRedshiftEvolution,arcus2021FastRadioBurst}. Then, by considering the $1+z$ factor in the observed host galaxy DM, the redshift of the FRBs can be solved numerically with \texttt{SciPy} \citep{virtanen2020SciPyFundamentalAlgorithms}. We acknowledge that there is a large uncertainty in estimating $z$ but nonetheless adopt the solution as the best estimate of the $z$ value of a particular burst. We also set a minimum redshift of $0.00225$, corresponding to a luminosity distance of $\SI{10}{\mega\parsec}$ to avoid zero or negative values.
    
    With redshift, we can derive several other features (as discussed below). We do not directly include redshift in our models. Rather, redshift is incorporated indirectly through other features.
    
    \item Brightness temperature (\si{\kelvin})

    For an FRB with peak specific flux $S_\nu$, duration $\Delta t$ and observed central frequency $\nu$, and redshift $z$, the full expression of brightness temperature should be:
    \begin{align}
        T_B&=\frac{S_\nu D_{\rm A}^2} {2\upi k_B (\nu \Delta t)^2} (1+z)^3 = \frac{S_\nu D_{\rm L}^2} {2\upi k_B (\nu \Delta t)^2 (1+z)} \nonumber \\
        &=\SI{1.1e35}{\kelvin} \qty(\frac{S_{\nu}}{\si{\jansky}}) \qty(\frac{\nu}{\si{\giga\hertz}})^{-2} \qty(\frac{\Delta t}{\si{\milli\s}})^{-2} \qty(\frac{D_{\rm A}}{\si{\giga\parsec}})^{2}(1+z)^3 \nonumber \\
        &=\SI{1.1e35}{\kelvin} \qty(\frac{S_{\nu}}{\si{\jansky}}) \qty(\frac{\nu}{\si{\giga\hertz}})^{-2} \qty(\frac{\Delta t}{\si{\milli\s}})^{-2}\qty(\frac{D_{\rm L}}{\si{\giga\parsec}})^{2}\frac{1}{1+z}.
    \label{eq:TB}
    \end{align}
    where $k_B$ is Boltzmann constant, $D_A$ is the angular diameter distance of the FRB, $D_L$ is the luminosity distance of the FRB.
    
    This expression differs from the previously suggested expression involving the angular distance $D_{\rm A}$ \citep{zhang2020PhysicalMechanismsFast, xiao2021PhysicsFastRadio, xiao2022NewInsightsCriterion} by including a $(1+z)^3$ correction factor to account for cosmological effects. For a flat universe, the luminosity distance is connected with the angular distance via $D_{\rm L}=D_{\rm A}(1+z)^2$ \citep{etherington1933LXDefinitionDistance,ellis2007DefinitionDistanceGeneral}, so $T_B$ can be also calculated with $D_{\rm L}$ with a $(1+z)^{-1}$ correction factor. There are two independent ways of deriving Equation (\ref{eq:TB}) in a self-consistent way, which are presented in Appendix \ref{sec:derivation_T_B}.

    \item Rest-frame width (\si{\s})
    
    The rest-frame width is calculated by correcting the time-dilation effect, i.e.
    \begin{equation}
        \Delta t_r = \frac{\Delta t}{1+z},
    \end{equation}
    where $\Delta t$ is the observed sub-burst width given by \texttt{fitburst}. This value is fitted with an FRB profile model and is different for different sub-bursts in the same burst. 

    \item Rest-frame frequency bandwidth (\si{\mega\hertz})
    \begin{equation}
        \Delta \nu=(\nu_{\rm max}-\nu_{\rm min})(1+z)
    \end{equation}
    where $\nu_{\rm max}$ and $\nu_{\rm min}$ are the highest and lowest observed frequencies of the burst reported in the CHIME catalog in units of $\si{\mega\hertz}$.
    
    \item Burst energy (\si{\erg})
    
    We follow \citet{zhang2018FastRadioBurst,zhang2021EnergyRedshiftDistributions,li2021BimodalBurstEnergy,hashimoto2022EnergyFunctionsFast} to calculate the burst energy:
    \begin{equation}
        E=4\upi D_{\rm L}^2 F \nu_c /(1+z)
    \end{equation}
    Where $F$ is the fluence of the FRB, and $\nu_c$ is peak frequency. Adopting $\nu_c$ rather than the bandwidth is more appropriate for wide-spectra FRBs such as the majority of the non-repeating ones \citep{zhang2018FastRadioBurst}. It overestimates the energy of FRBs if the spectra have narrow bands. For consistency, we adopt $\nu_c$ for all the bursts since the majority of the bursts are apparently non-repeating ones.
    
\end{itemize}

\section{Methods}
\label{sec:methods}

\subsection{Pre-processing}
\label{subsec:pre-processing}

We randomly split our data in a 7:3 ratio into a training set and a test set while keeping the original ratio of repeating to non-repeating FRBs in both sets. The former is used to train the machine learning algorithms, and the latter is used to test the performance of the machine learning models in classifying the repeating and non-repeating FRBs. This split gives us 415 bursts (66 repeating and 349 non-repeating) in the training set and 179 bursts (28 repeating and 151 non-repeating) in the test set.

The input features have different ranges. The brightness temperature and burst energy are at orders $\sim10^{30}$ and span across several orders of magnitude, while the box-car width and rest-frame width are at orders of $\sim10^{-2}$. Hence, we need to normalize the input features. We first take the base 10 logarithms of brightness temperature, burst energy, and frequency bandwidth. We also convert the box-car width and rest-frame width to milliseconds. Then, we standardize the input features by removing the mean and scaling the variance to unity. The standard distribution scaler is trained with the training set only, and subsequently applied to all the data.

The distribution of the two types of FRBs in our sample is significantly imbalanced. The size of the non-repeating FRB sample is $\sim5\times$ that of the repeating FRBs. Because the true ratio of repeating to non-repeating FRBs is unknown, we need to avoid our models learning this apparent ratio. Therefore, we use the synthetic minority over-sampling technique (SMOTE) \citep{chawla2002SMOTESyntheticMinority} as implemented in \texttt{imbalanced-learn} \citep{lemaitre2017ImbalancedlearnPythonToolbox} to over-sample the repeating FRB training set and make it the same size as the non-repeating FRB learning sample. Note that we only over-sample the training set, but keep the imbalance in the test set. This way we do not introduce any augmented data into the testing process \citep{santos2018CrossValidationImbalancedDatasets}. After data augmentation, there are the same number of repeating and non-repeating FRBs in the training set (349).

\subsection{Model evaluation}
\label{subsec:model evaluation}

After training the models with the training set, we test the models with the test set. The most intuitive model evaluation metric is accuracy, defined as the ratio between the number of correctly predicted labels to the total number of input data points. However, this simple metric cannot be applied to our imbalanced data. Even a model simply predicts every FRB to be non-repeating can score 84.26\% accuracy with our test set. Therefore, we need to find a more reliable method to evaluate our models.

There are two commonly used metrics in evaluating the performance of machine learning models, precision and recall:
\begin{itemize}
    \item Precision
    \begin{equation}
        \mathrm{Precision} = \frac{\mathrm{True\;positives}}{\mathrm{True\;positives} + \mathrm{False\;positives}}
    \end{equation}
    Precision measures how many of the items predicted by the model as positive (in this study repeating FRBs) are true positives.
    
    \item Recall
    \begin{equation}
        \mathrm{Recall} = \frac{\mathrm{True\;positives}}{\mathrm{True\;positives} + \mathrm{False\;negatives}}
    \end{equation}
    Recall measures how many of the originally positive items are correctly predicted as positive by the model.
\end{itemize}

These two metrics are then usually combined to form the $F_{\beta}$ score \citep{vanrijsbergen1979InformationRetrieval,sasaki2007TruthFmeasure}:
\begin{equation}
    F_{\beta} = (1+\beta^2)\frac{\mathrm{Precision}\cdot\mathrm{Recall}}{(\beta^2\cdot\mathrm{Precision})+\mathrm{Recall}}
\end{equation}
Where $\beta$ is a weighting factor for precision and recall. The $F_{\beta}$ score have a range of $0\sim1$, with $F_{\beta}=1$ signaling a perfect model.

While it is common practice to let $\beta=1$ and weight precision and recall equally, in this study precision and recall are not equal. An apparently non-repeating FRB could be a hidden repeating FRB yet to be seen repeating, but a repeating FRB cannot be a misidentified non-repeating one. A model that produces some false positives while yielding few false negatives should not be rejected. Therefore we let $\beta=2$ to prioritize recall over precision. This way we can select models that find most apparent repeating FRB successfully, while retaining their ability to uncover possible hidden repeaters. We will refer to this metric as the $F_2$ score for the rest of the paper.

\section{Results}
\label{sec:results}

\subsection{Simple decision tree method}
\label{subsec:simple_decision_tree}

Classification and Regression Trees (CART) is a type of simple machine learning algorithms where a decision tree of simple decision rules is created based on input features to predict corresponding labels or output values of new data \citep[e.g.][]{breiman1984ClassificationRegressionTrees,timofeev2004ClassificationRegressionTrees,loh2011ClassificationRegressionTrees,loh2014FiftyYearsClassification}. In this study, we build classification decision trees with the input features described in Section \ref{sec:data} and methods described in Section \ref{sec:methods}.

We found the default hyper-parameter settings of \texttt{scikit-learn} mostly work well for our problem, and altering the hyper-parameters does not really improve the results. Thus for the rest of this paper, we use default hyper-parameters unless otherwise stated. We also list the hyper-parameters used in each method in Table \ref{table:paramters_list}.

\begin{figure*}
    \centering
	\includegraphics[width=0.99\textwidth]{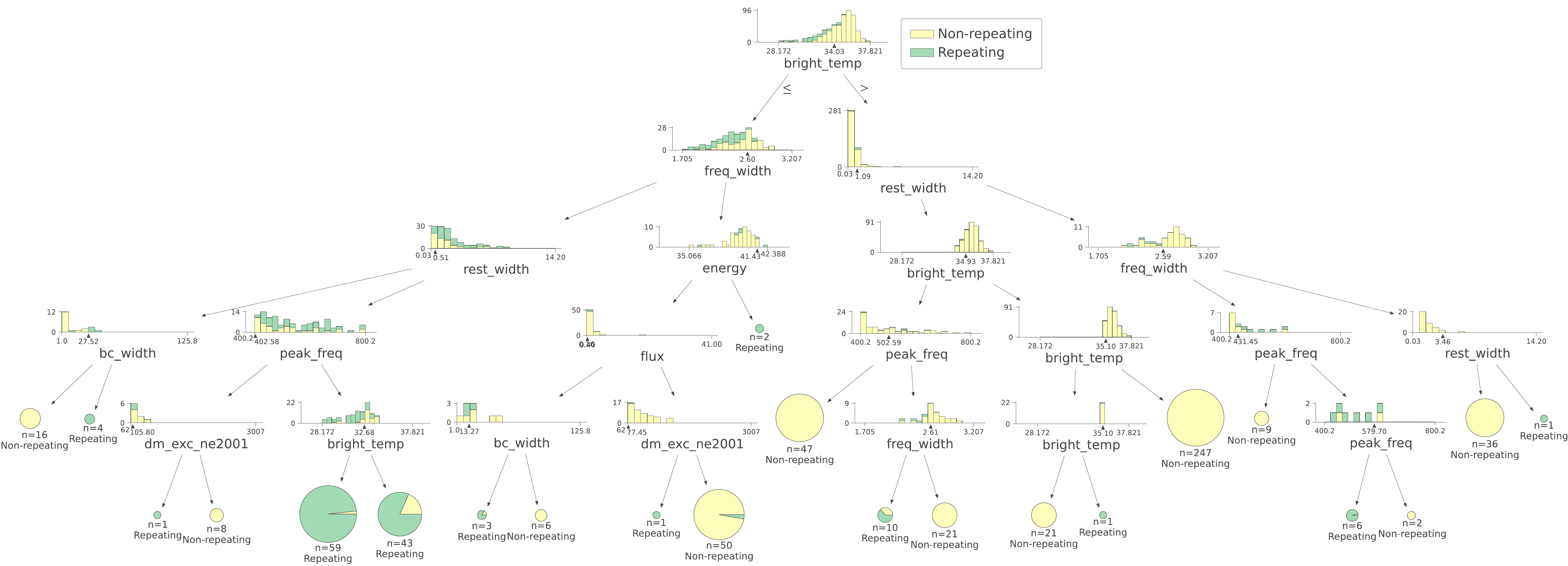}
	\caption{Decision paths of the decision tree on the entire dataset.}
	\label{fig:tree_vis}
\end{figure*}

We build a decision tree with a max depth of 5, and we are able to achieve an average $F_2$ score of 0.7351. The decision path of one trial of this tree method is shown in Fig. \ref{fig:tree_vis}. An example of the confusion matrix of this method is shown in Fig. \ref{fig:tree_cm}.

We then build another decision tree with a max depth of 3, and only two input features: brightness temperature and rest-frame frequency bandwidth. We choose these two features because they are found to have the most feature importance later in this section. For this decision tree with 2 features, we are able to achieve average $F_2$=0.7369. An example of the confusion matrix is shown in Fig. \ref{fig:2d_cm}.

\begin{figure}
	\includegraphics[width=0.49\textwidth]{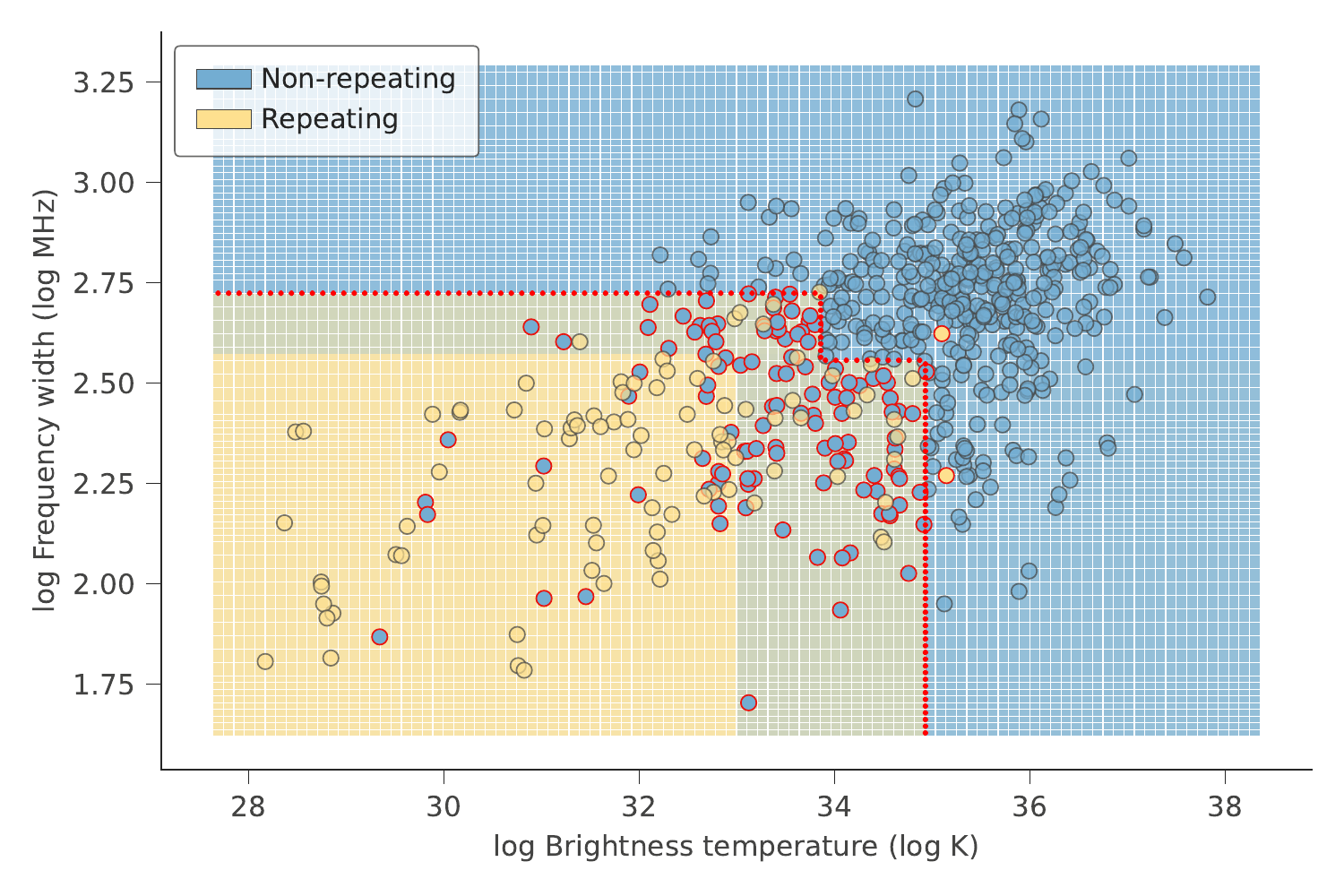}
	\caption{Decision boundary of the decision tree with 2 features on the entire dataset. The depth of colors represent the probability of either class.}
	\label{fig:2d_boundary}
\end{figure}

Fig. \ref{fig:2d_boundary} shows the log-log plot of brightness temperature and intrinsic frequency width. Most non-repeating FRBs reside in the top right corner, whereas most repeating FRBs reside in the lower left corner. While the repeating FRBs generally stay in their territory, many non-repeating FRBs permeate into the repeating region. This on one hand shows that the parameter distributions of repeating and non-repeating FRBs are significantly different, and on the other hand that some repeating FRBs are likely mis-classified as non-repeating.

\begin{figure*}
\centering
\begin{subfigure}{0.49\textwidth}
	\includegraphics[width=\textwidth]{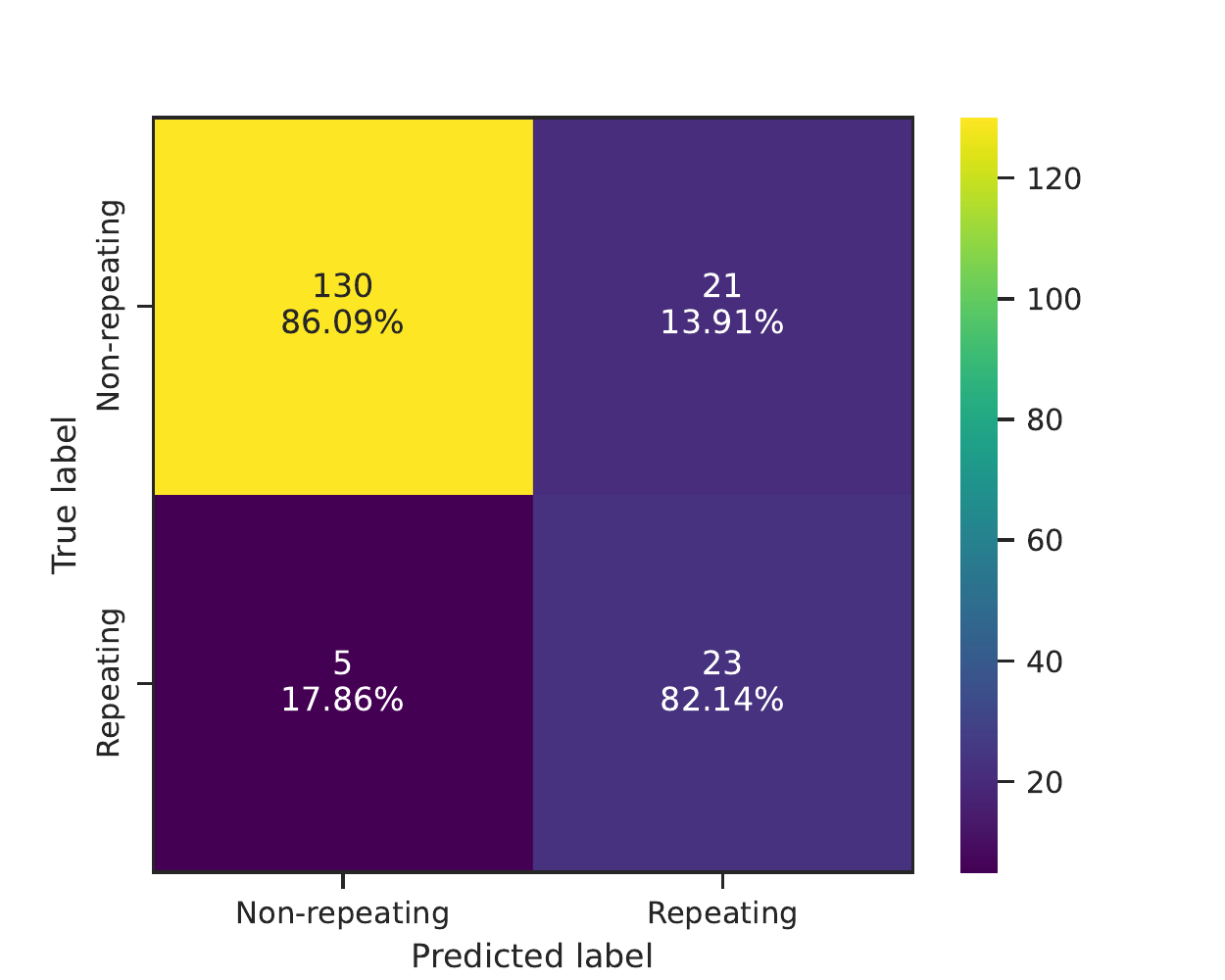}
	\caption{Decision tree}
    \label{fig:tree_cm}
\end{subfigure}
\begin{subfigure}{0.49\textwidth}
	\includegraphics[width=\textwidth]{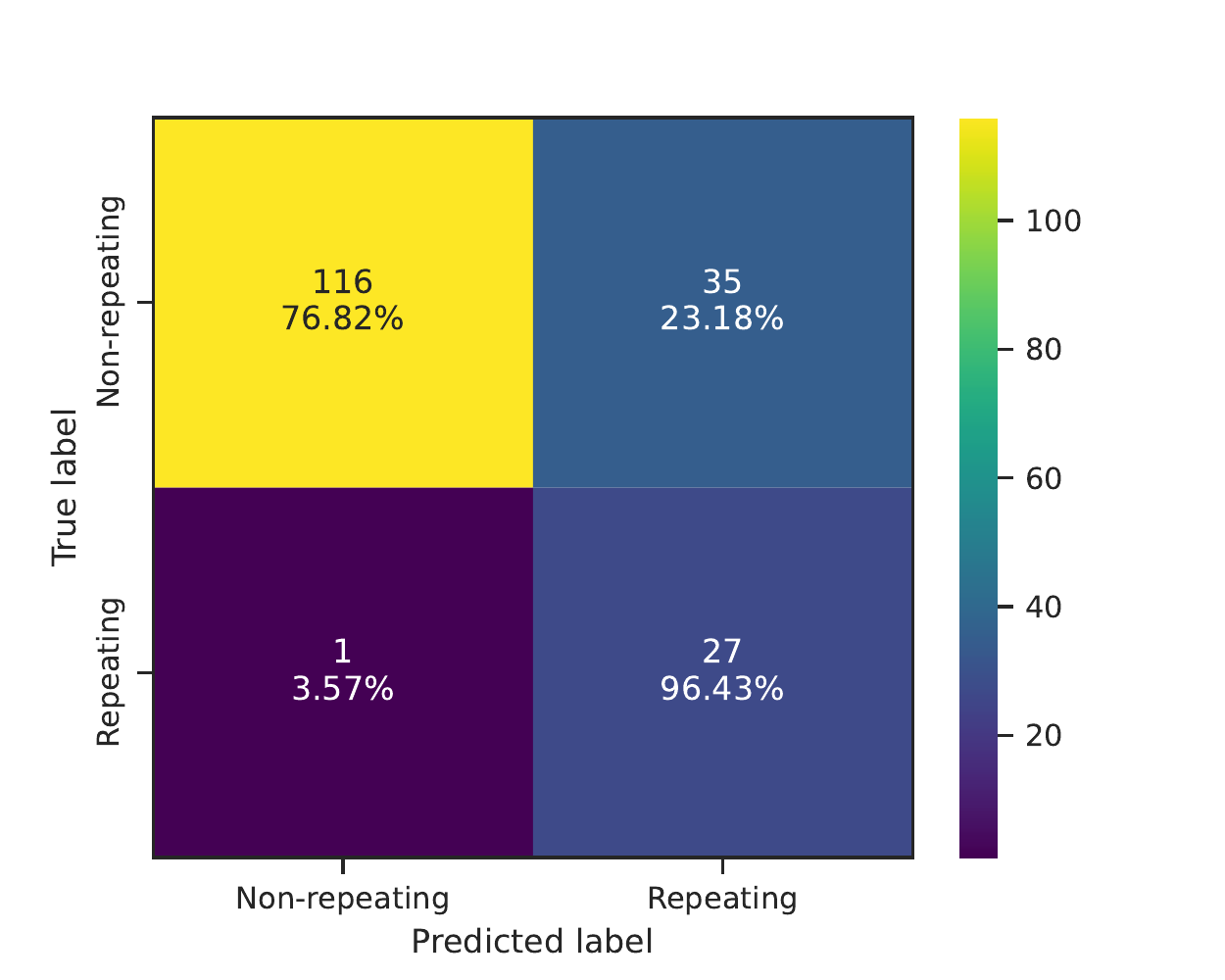}
	\caption{Decision tree with 2 features}
    \label{fig:2d_cm}
\end{subfigure}
\begin{subfigure}{0.49\textwidth}
	\includegraphics[width=\textwidth]{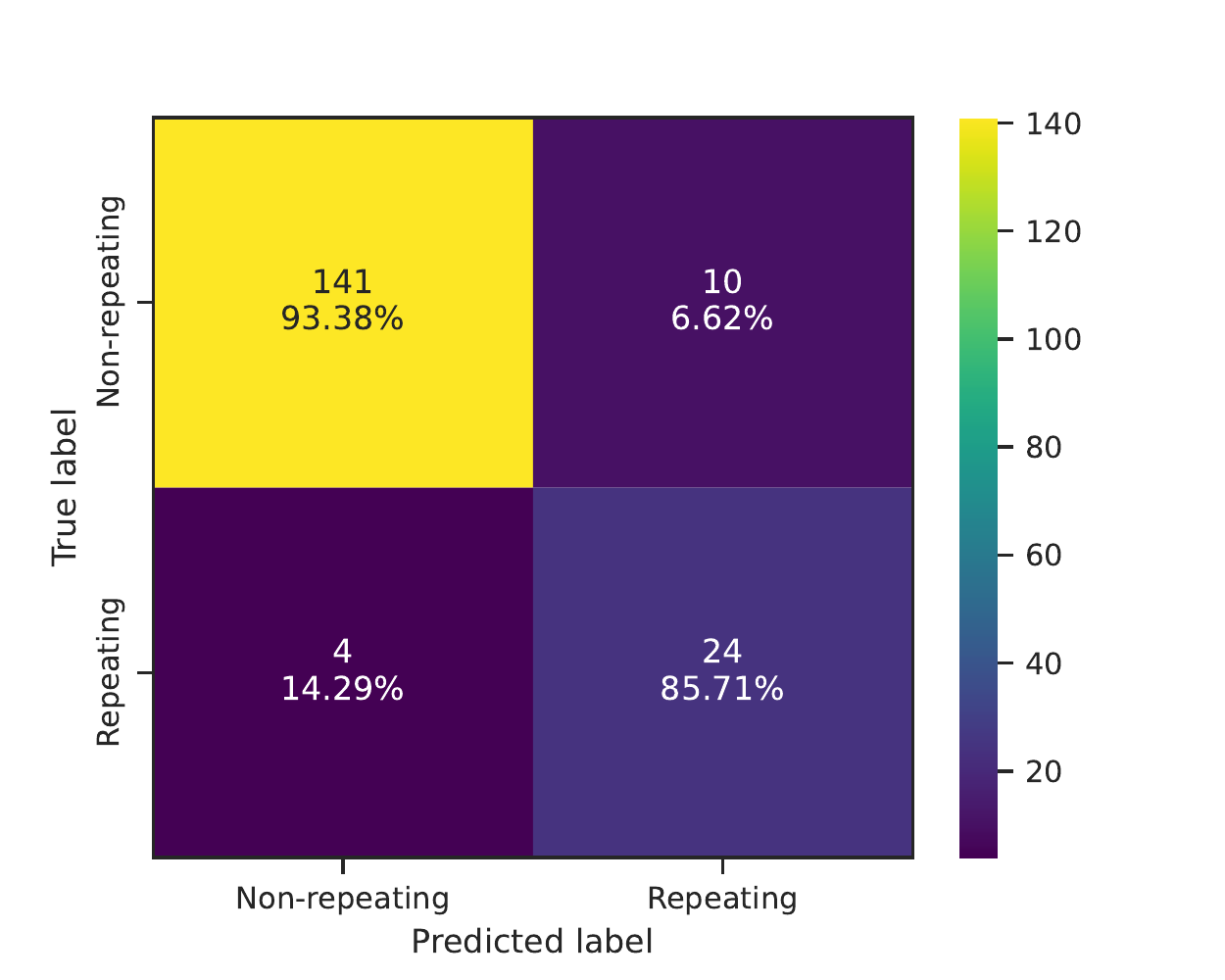}
	\caption{Random forest}
    \label{fig:rf_cm}
\end{subfigure}
\begin{subfigure}{0.49\textwidth}
	\includegraphics[width=\textwidth]{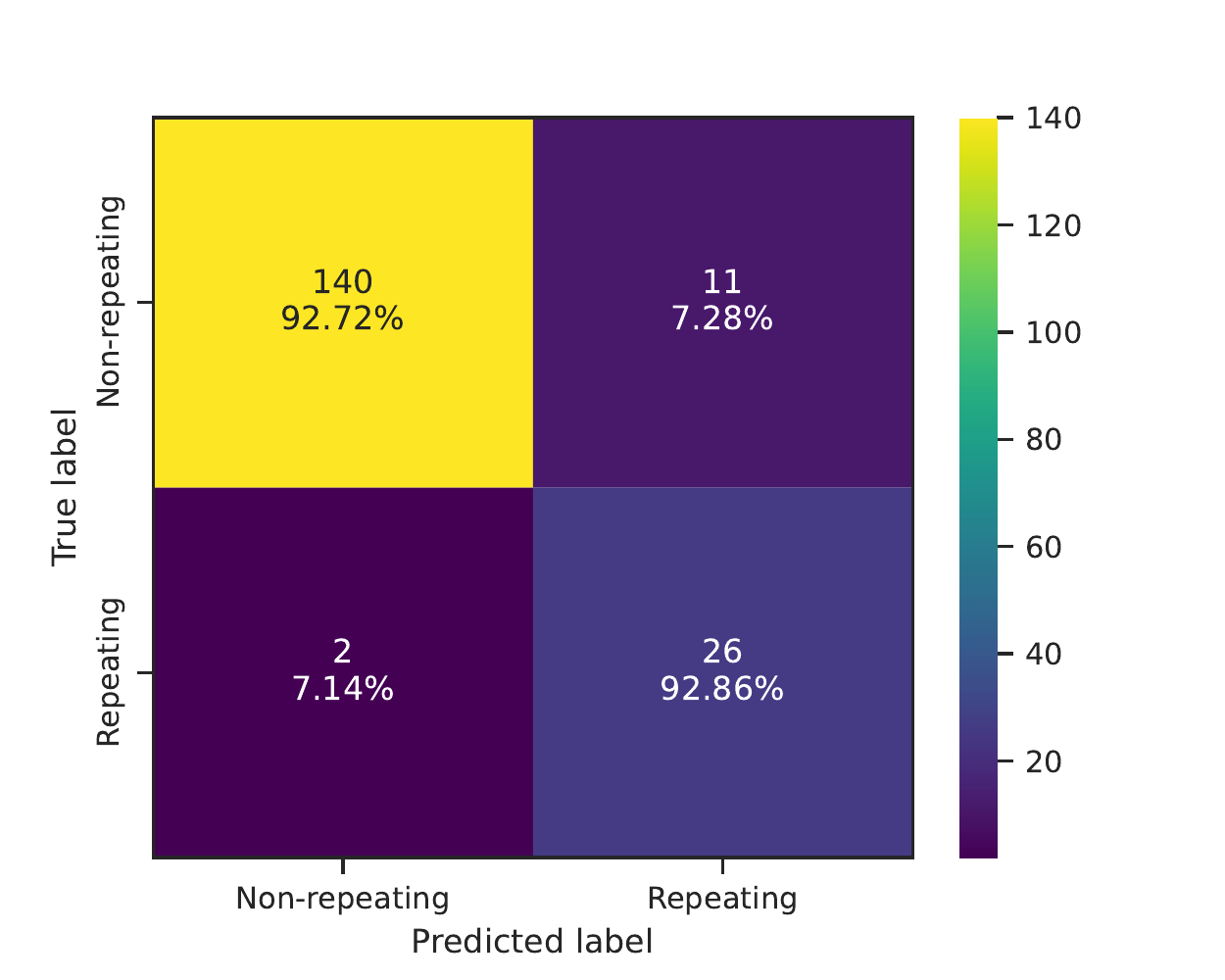}
	\caption{AdaBoost}
    \label{fig:ab_cm}
\end{subfigure}
\begin{subfigure}{0.49\textwidth}
	\includegraphics[width=\textwidth]{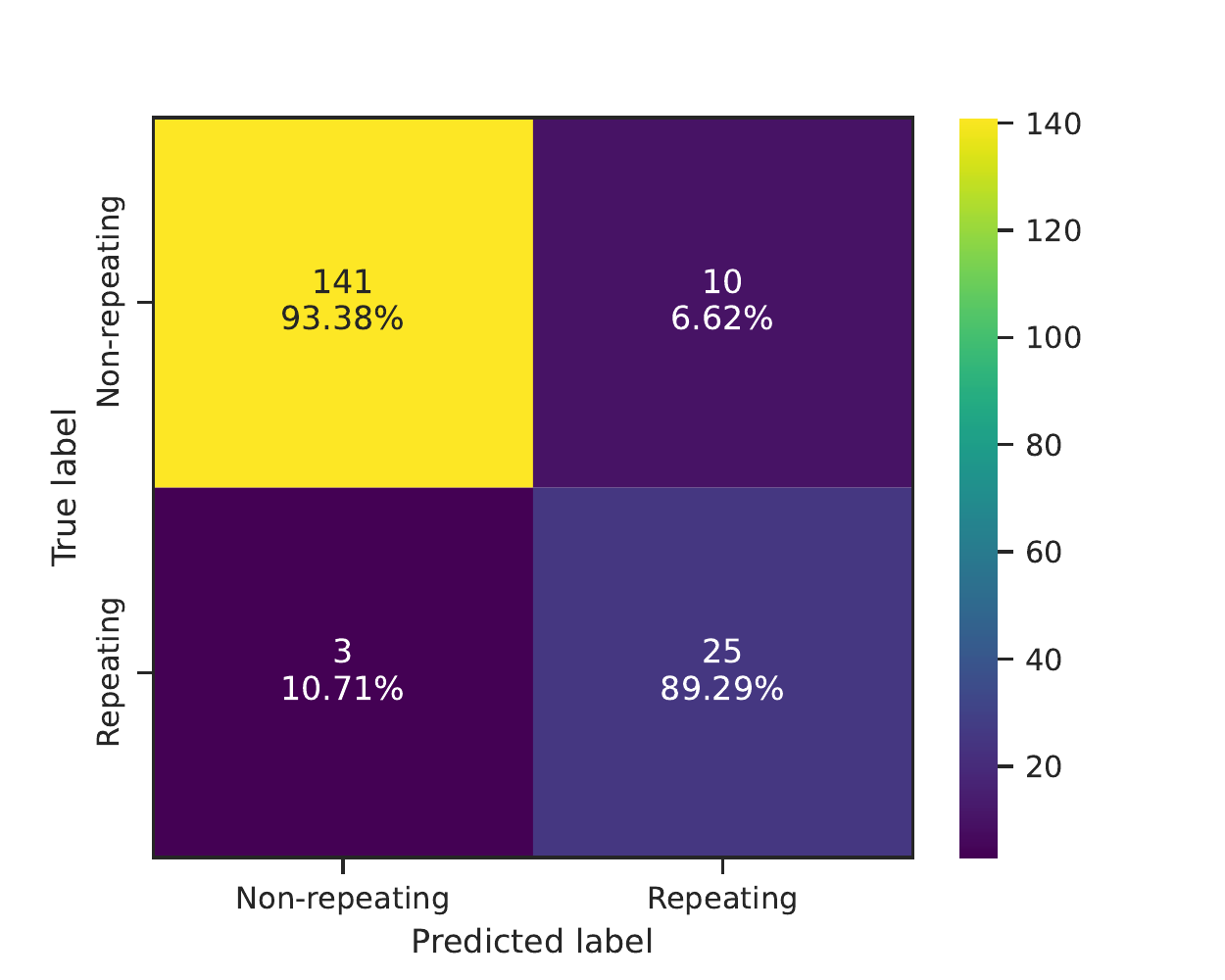}
	\caption{LightGBM}
    \label{fig:lgbm_cm}
\end{subfigure}
\begin{subfigure}{0.49\textwidth}
	\includegraphics[width=\textwidth]{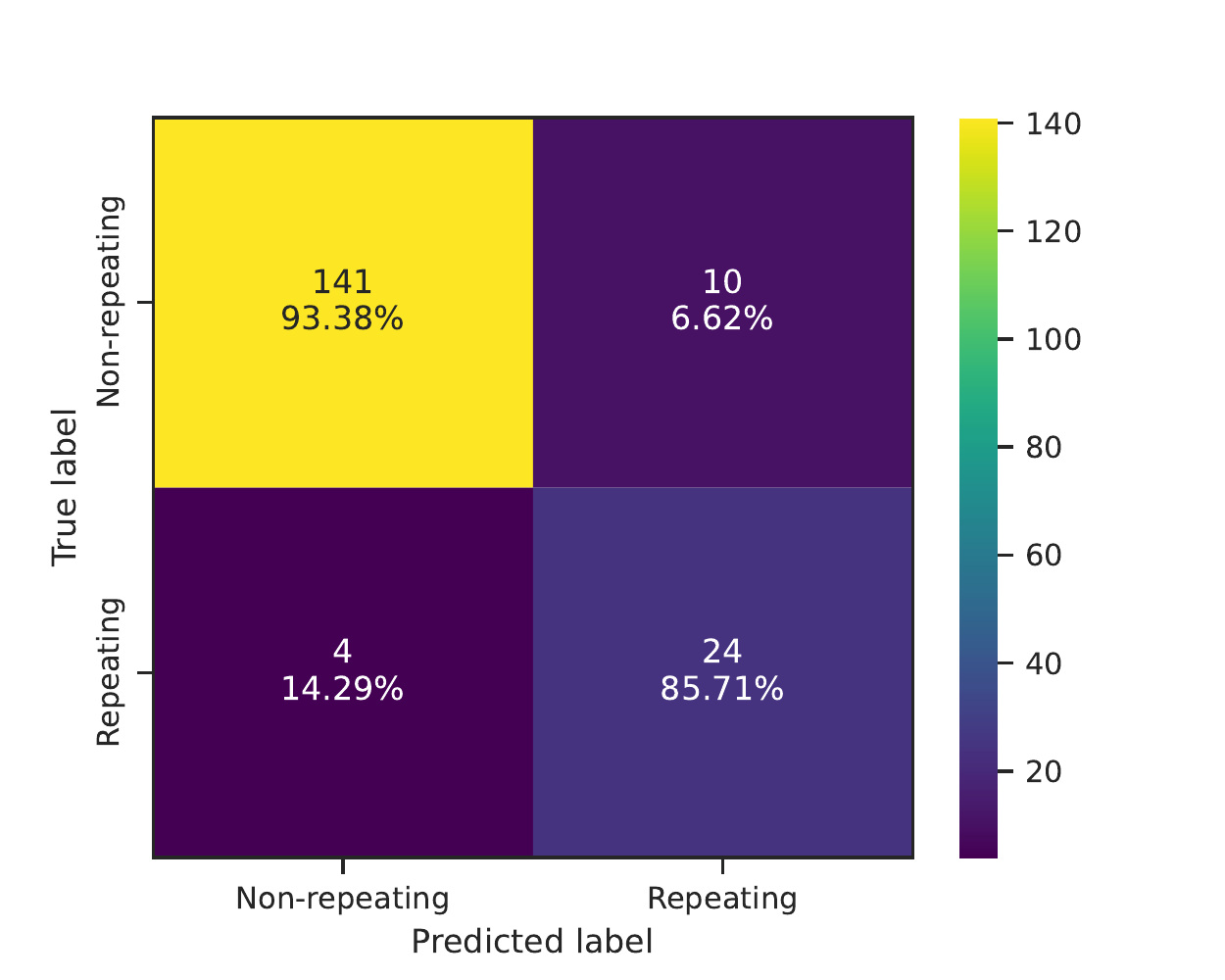}
	\caption{XGBoost}
    \label{fig:xgb_cm}
\end{subfigure}
\caption{Examples of confusion matrices of different methods on the test set.}
\label{fig:cm}
\end{figure*}%
\begin{figure*}\ContinuedFloat
\centering
\begin{subfigure}{0.49\textwidth}
	\includegraphics[width=\textwidth]{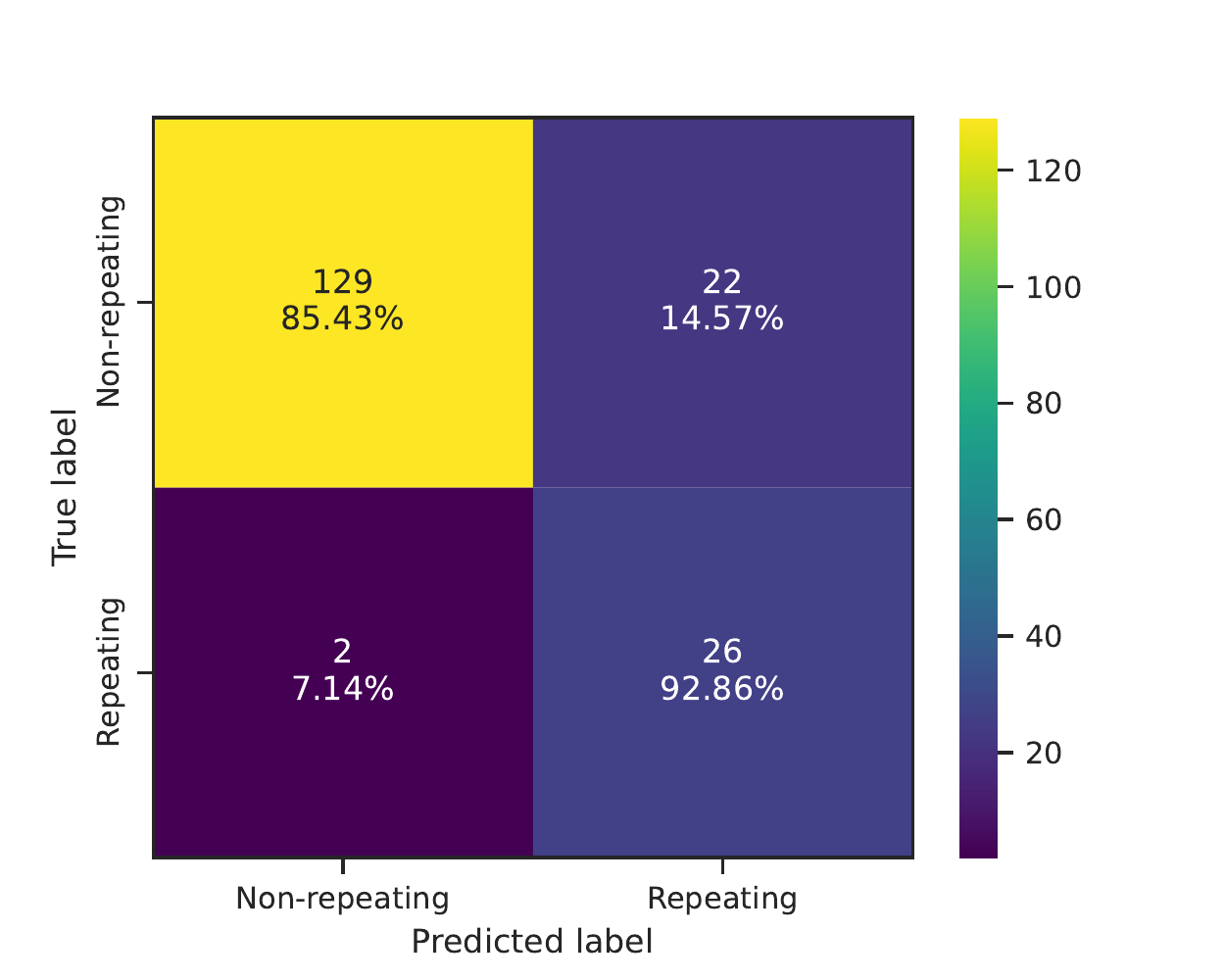}
	\caption{SVM}
    \label{fig:svm_cm}
\end{subfigure}%
\begin{subfigure}{0.49\textwidth}
	\includegraphics[width=\textwidth]{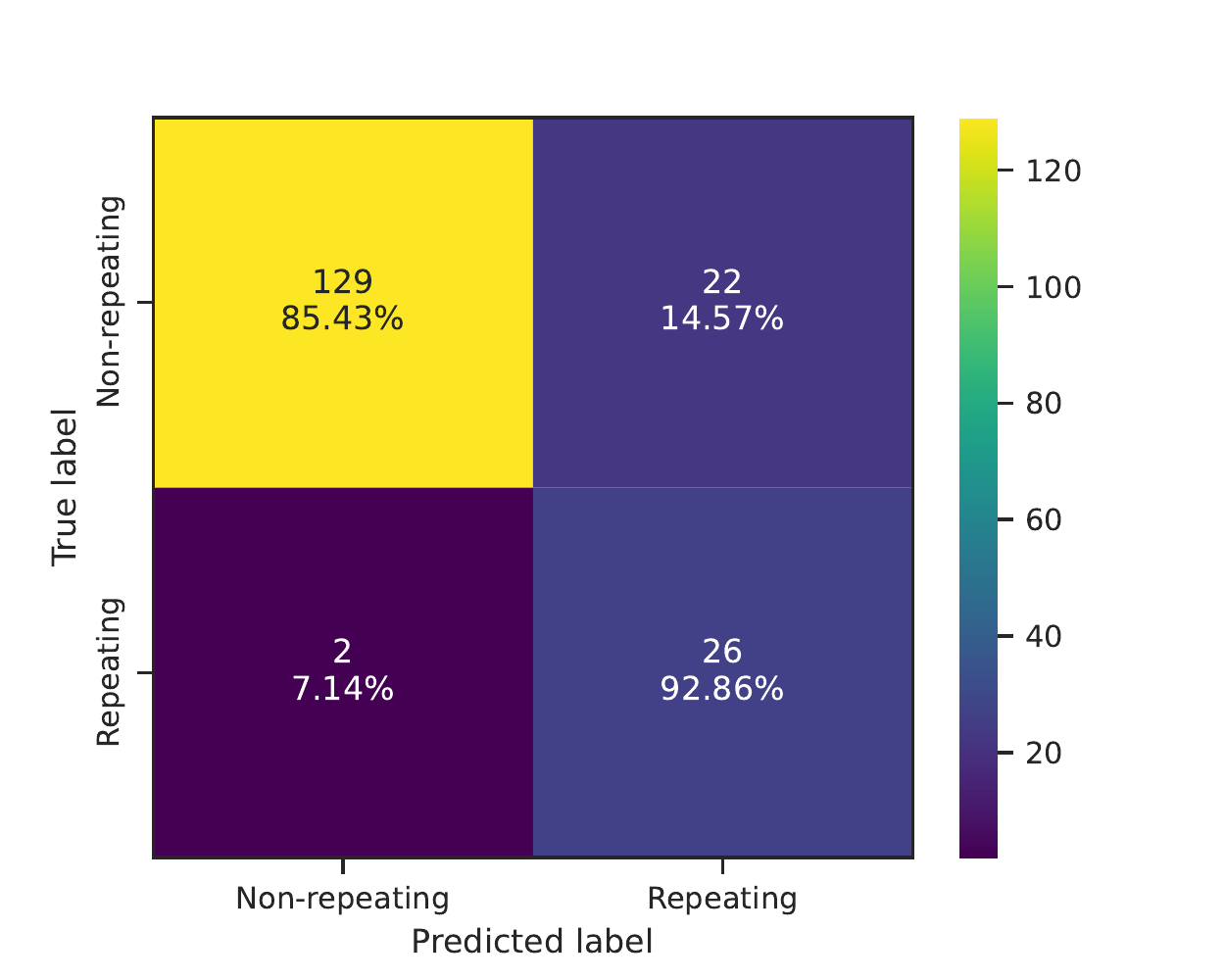}
	\caption{Nearest centroid}
    \label{fig:nc_cm}
\end{subfigure}
\caption{Examples of confusion matrices of different methods on the test set (cont.).}
\end{figure*}

\begin{figure*}
\centering
\begin{subfigure}{0.49\textwidth}
	\includegraphics[width=\textwidth]{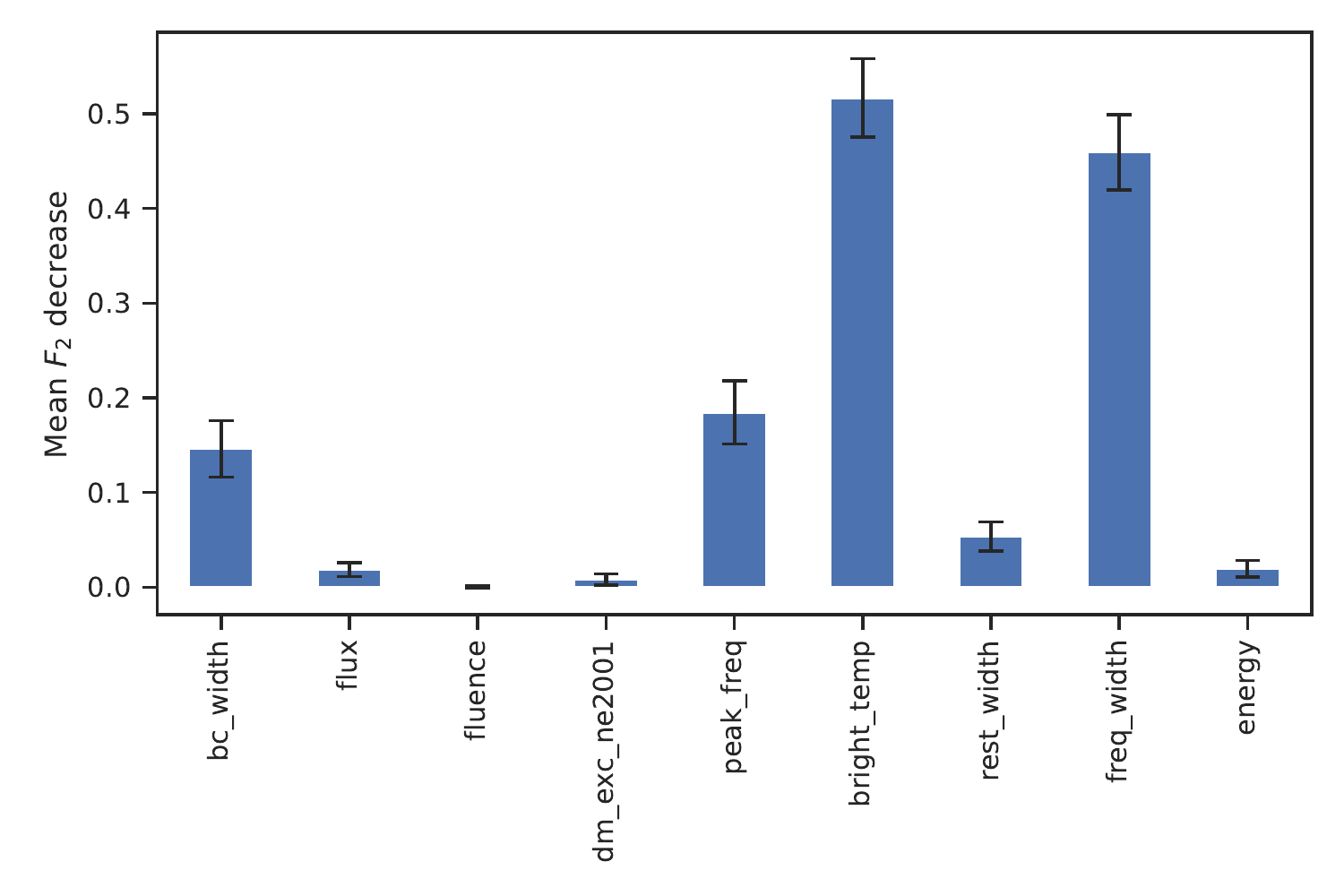}
	\caption{Random forest}
	\label{fig:rf_fi}
\end{subfigure}
\begin{subfigure}{0.49\textwidth}
	\includegraphics[width=\textwidth]{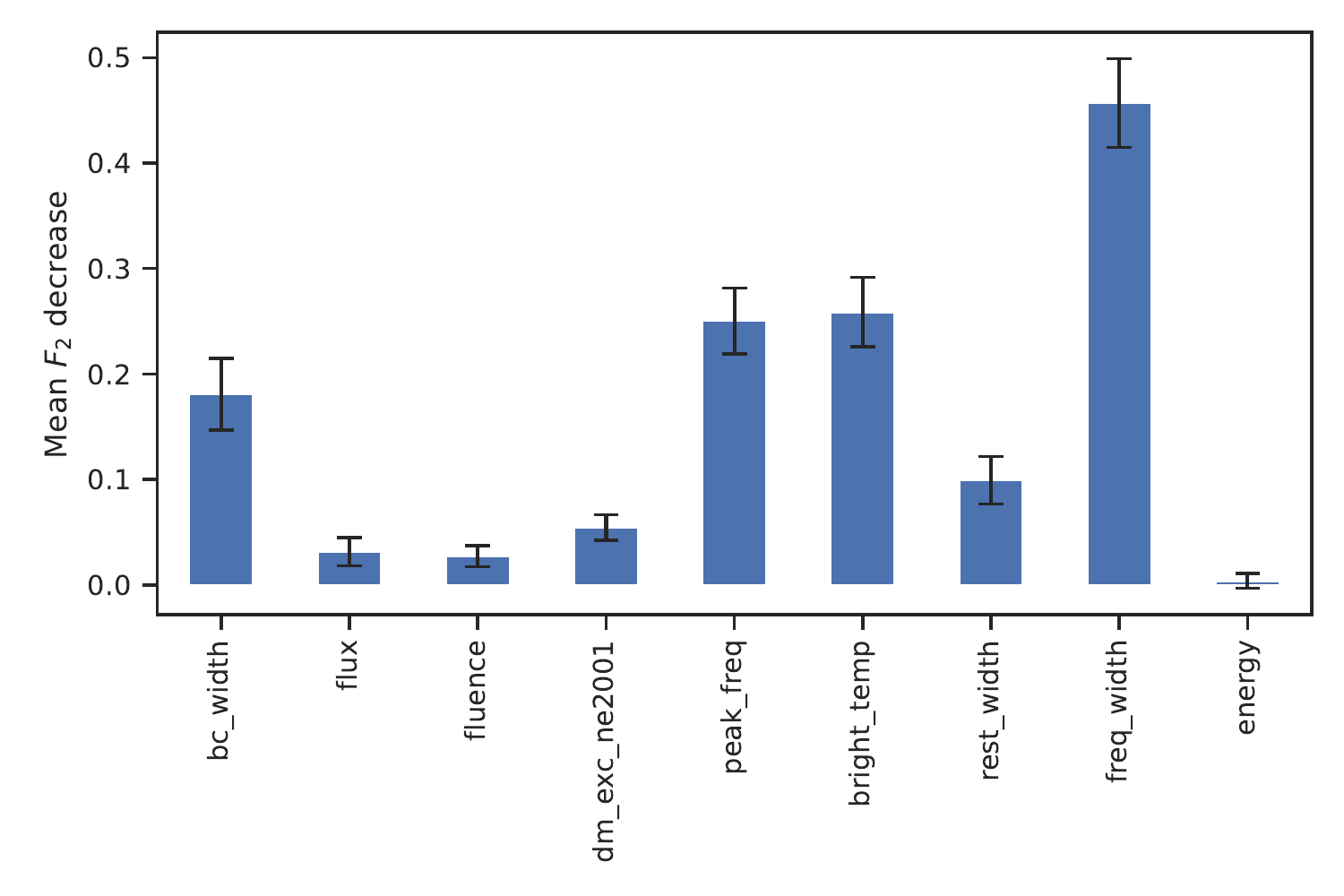}
	\caption{AdaBoost}
    \label{fig:ab_fi}
\end{subfigure}
\begin{subfigure}{0.49\textwidth}
	\includegraphics[width=\textwidth]{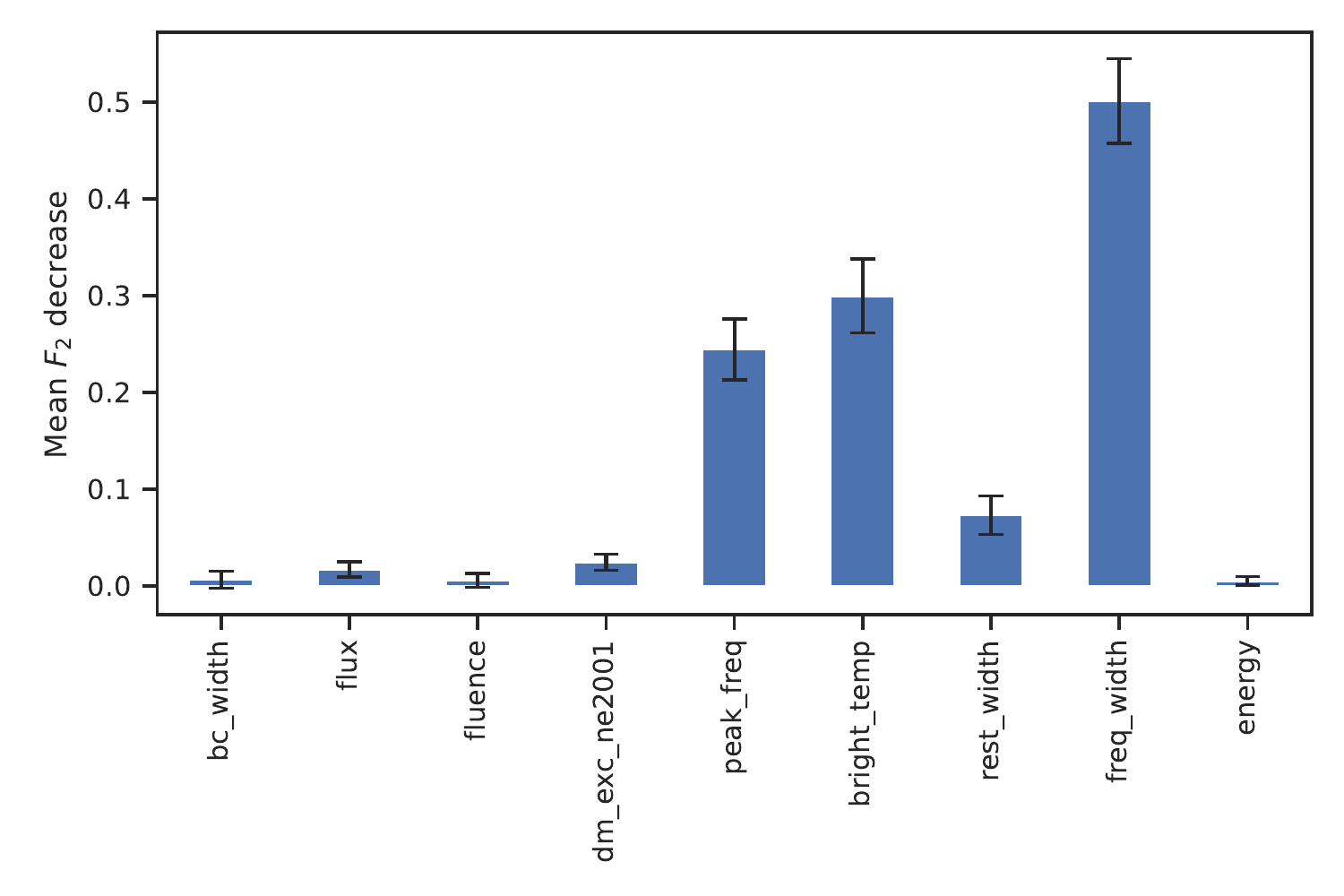}
	\caption{LightGBM}
    \label{fig:lgbm_fi}
\end{subfigure}
\begin{subfigure}{0.49\textwidth}
	\includegraphics[width=\textwidth]{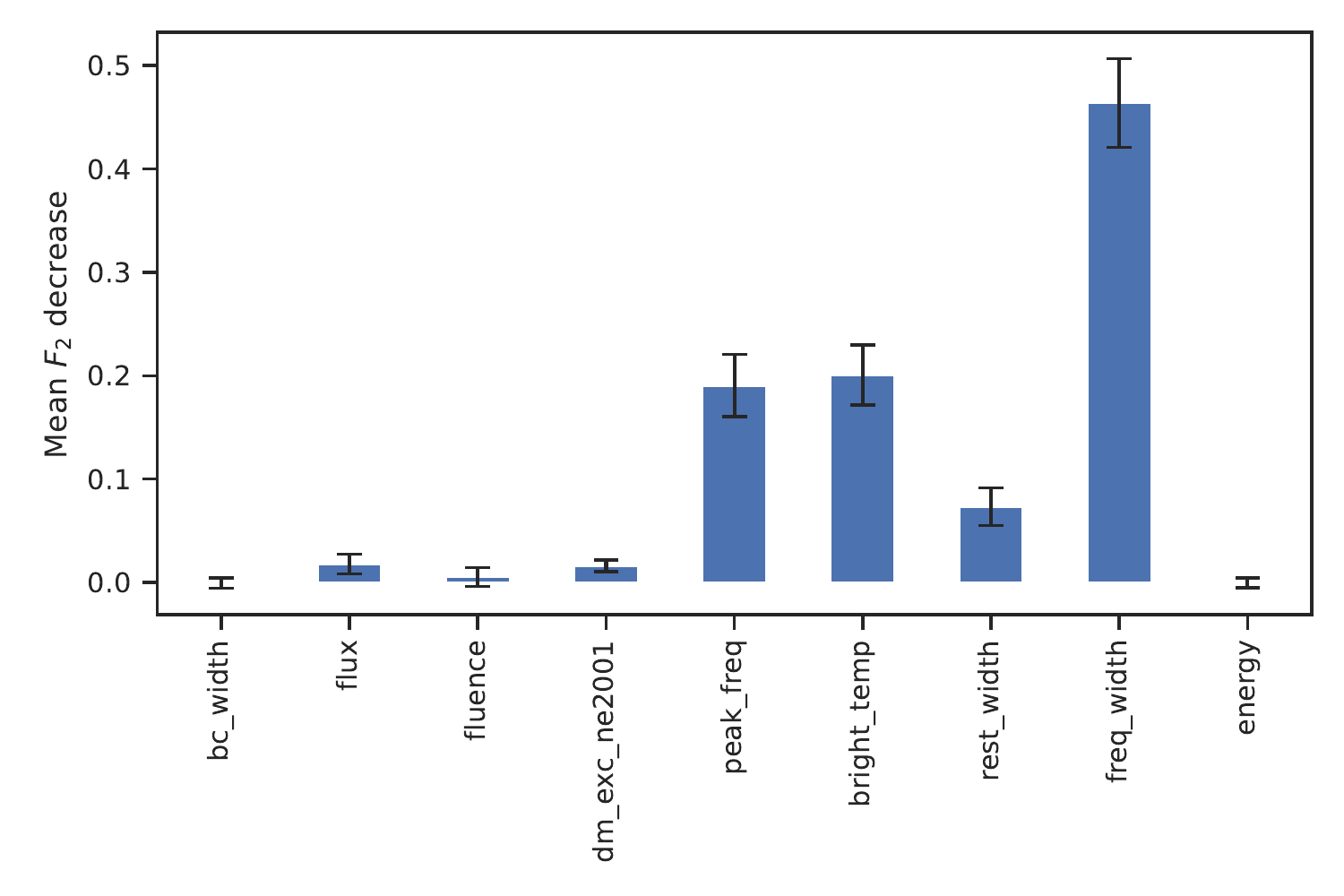}
	\caption{XGBoost}
    \label{fig:xgb_fi}
\end{subfigure}
\begin{subfigure}{0.49\textwidth}
	\includegraphics[width=\textwidth]{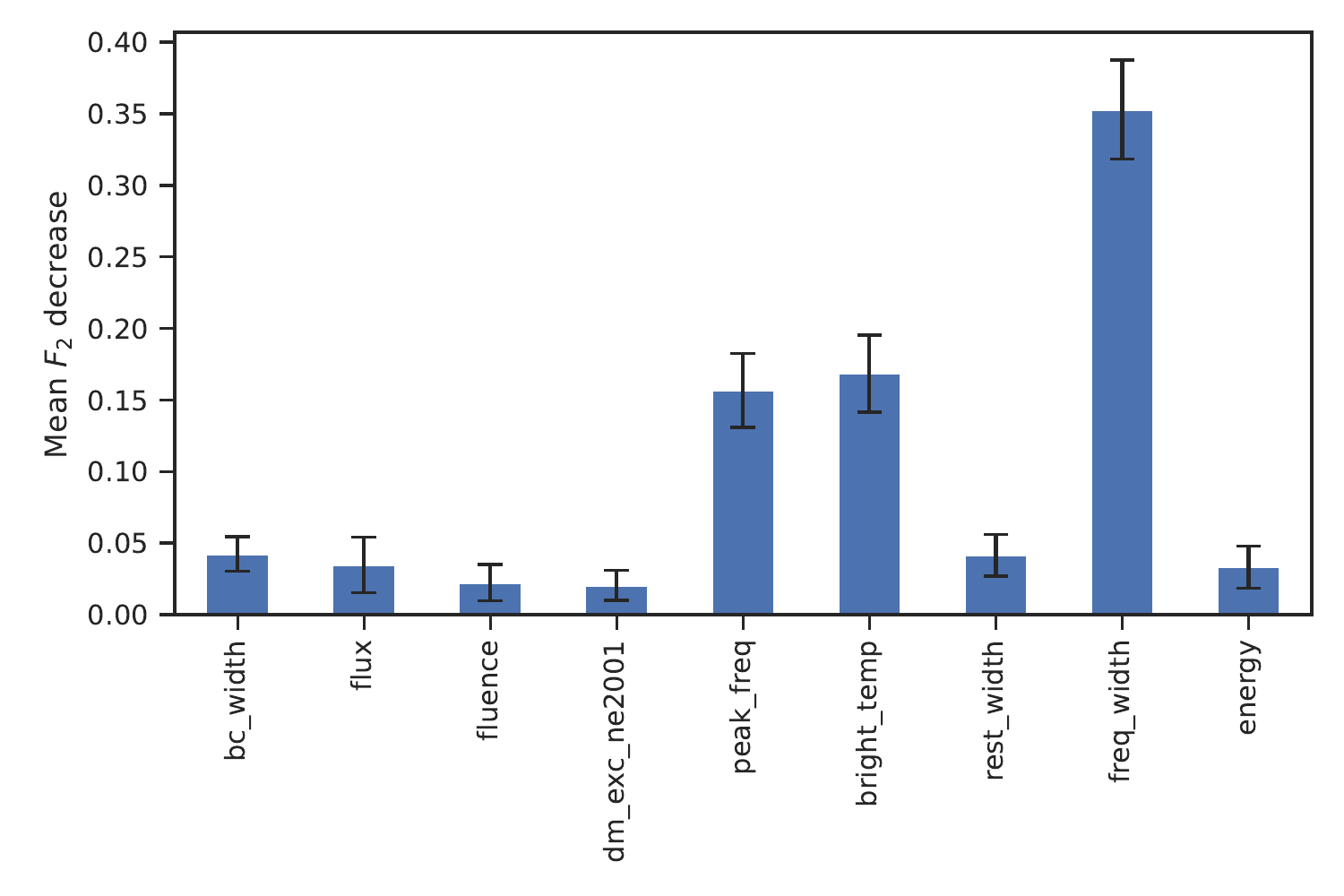}
	\caption{SVM}
    \label{fig:svm_fi}
\end{subfigure}
\begin{subfigure}{0.49\textwidth}
	\includegraphics[width=\textwidth]{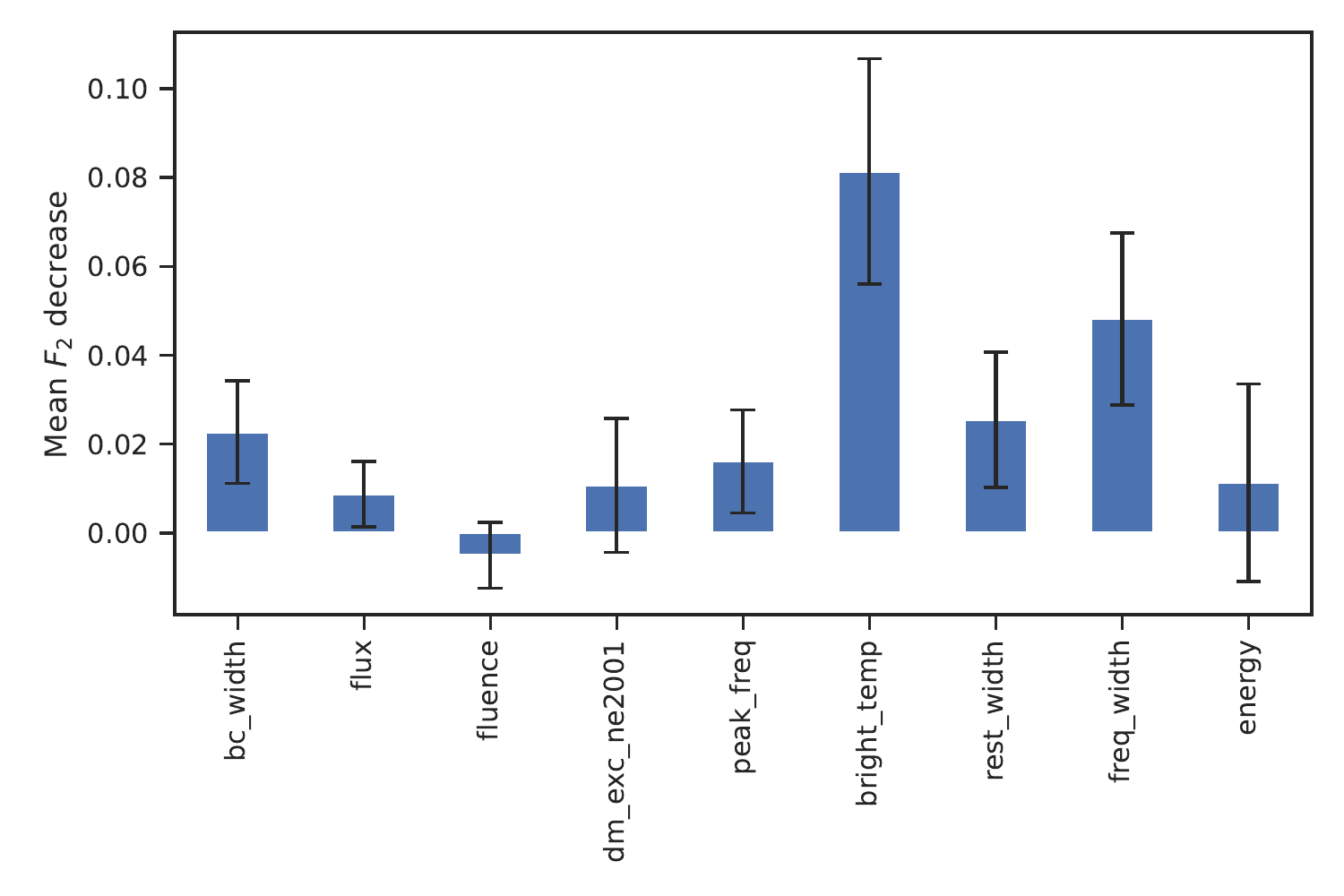}
	\caption{Nearest centroid}
    \label{fig:nc_fi}
\end{subfigure}
\caption{Permutation feature importance of different methods on the entire dataset.}
\label{fig:fi}
\end{figure*}

\subsection{Random forest method}
\label{subsec:random_forest}

In machine learning, bias–-variance tradeoff is a crucial topic \citep[e.g.][]{geman1992NeuralNetworksBias,friedman1997BiasVarianceLoss,neal2019ModernTakeBiasVariance}. Bias measures the difference between the predictions from the models to the training input, while variance measures how accurately the model can predict with new data not present in the training process. Since all models would eventually be used to predict newly obtained data, variance is just as important as bias. While a model with both low bias and low variance is preferable, this is essentially not possible in reality. Therefore, one must try to balance bias and variance, so as to get accurate model predictions for both the training set and the test set or future data.

For a single decision tree, it is difficult to balance bias and variance. If the depth is too shallow, then the decision tree will have difficulty in detecting all the positive items. On the other hand, if the depth is too deep, the decision tree will over-fit on the training data and perform badly on the test data.

To solve this problem, we employ the random forest algorithm \citep{breiman2001RandomForests}. We randomly select sub-samples of the training set, and train many deep decision trees with the random sub-samples that tend to over-fit. This process is called bagging \citep{breiman1996BaggingPredictors}. Then, the prediction of the ensemble of the deep decision trees is taken average to generate the output of the entire model to mitigate the over-fitting problem.

With this method, we are able to obtain an average $F_2$ score of 0.7821. An example of the confusion matrix is shown in Fig. \ref{fig:rf_cm}.

We also look into the importance of each input feature with permutation feature importance \citep{breiman2001RandomForests,altmann2010PermutationImportanceCorrected,fisher2019AllModelsAre} on the entire dataset. Each time the values of one input feature are randomly shuffled across all data points, removing the correlation between this feature and the output label. The drop in model metric is observed to denote the importance of the feature. This process is repeated 100 times for each feature and the average from the trials are used as the final feature importance. The error bars are drawn with standard deviations from the trials.

The permutation feature importance for the random forest models is shown in Fig. \ref{fig:rf_fi}. Again, brightness temperature and rest-frame frequency bandwidth are of the highest importance to this model.

\subsection{Boosted decision tree method}
\label{subsec:boosted_decision_tree}

Another way to incorporate many decision trees into a single predictor is through boosting. In contrast to random forest, which uses deep decision trees trained on part of the data, boosting uses shallow decision trees trained on the entire data. The weak predictors are again taken average to obtain the final output \citep{freund1997DecisionTheoreticGeneralizationOnLine,drucker1997ImprovingRegressorsUsing,friedman1998AdditiveLogisticRegression,schapire1999ImprovedBoostingAlgorithms,mayr2014EvolutionBoostingAlgorithms}.

There are different algorithm implementations of boosting. In this study, we employ AdaBoost \citep{freund1997DecisionTheoreticGeneralizationOnLine}, LightGBM \citep{ke2017LightGBMHighlyEfficient} and XGBoost \citep{chen2016XGBoostScalableTree}.

With AdaBoost, we obtain an average $F_2$ score of 0.7666. We also get average $F_2$ scores of 0.7832 and 0.7843 for LightGBM and XGBoost, respectively.

Examples of confusion matrices and feature importance are shown in Figs \ref{fig:ab_cm},\ref{fig:lgbm_cm},\ref{fig:xgb_cm} and Figs \ref{fig:ab_fi},\ref{fig:lgbm_fi},\ref{fig:xgb_fi}.

\subsection{Support vector machine method}
\label{subsec:SVM}

Support vector machine (SVM) \citep{cortes1995SupportVectorNetworks} projects the input features onto a hyper-space, and then attempts to find a hyper-plane to separate the data points based on input labels. Since most inputs are not linearly separable, studies most commonly use the "kernel trick" \citep{hofmann2008KernelMethodsMachine} to project the data points to another coordinate to make them linearly separable.

We are able to achieve an average $F_2$ score of 0.8180 with our SVM model. An example of the confusion matrix and feature importance are shown in Figs \ref{fig:svm_cm} and \ref{fig:svm_fi}.

\subsection{Nearest centroid method}
\label{subsec:nearest_centroid}

Similar to the SVM method, the nearest centroid method projects the data points into a hyperspace with their input features as position vectors \citep[e.g.][]{hastie2009ElementsStatisticalLearning}. The centroids of each class are learned by taking the average of the positions of each class of the learning set. The predictions for the test set are determined by the nearest class centroid. 

With this method, we can achieve an average $F_2$ score of 0.7147. An example of the confusion matrix and feature importance are shown in Figs \ref{fig:nc_cm} and \ref{fig:nc_fi}.

\subsection{Comparing different methods}
\label{subsec:compare_methods}

\begin{table}
\centering
    \begin{tabular}{ccc}
    \hline
    Model & $F_2$ Mean & $F_2$ SD \\
	\hline
	Decision tree (all features) & 0.7351 & 0.0632 \\
	Decision tree ($T_B$ and $\Delta\nu$) & 0.7369 & 0.0499 \\
	Random forest & 0.7821 & 0.0643 \\
	AdaBoost & 0.7666 & 0.0617 \\
	LightGBM & 0.7832 & 0.0647 \\
	XGBoost & 0.7843 & 0.0631 \\
	SVM & 0.8180 & 0.0483 \\
	Nearest centroid & 0.7147 & 0.0614 \\
	\hline
	\end{tabular}
	\caption{List of average $F_2$ scores and standard deviations obtained with different models.}
	\label{table:f2_list}
\end{table}

Because the training and test set split introduces randomness to the process, $F_2$ scores from single trial may not be able to fully reflect the prediction abilities of the machine learning models. Therefore, we repeat the random splitting and training process 1000 times, and record the $F_2$ scores of each model on the test set.

We report the average $F_2$ scores, along with standard deviations based on the 1000 trials for each model in Table \ref{table:f2_list}. We found that the SVM model performs the best in predicting repeating and non-repeating FRBs, while $F_2$ scores of all other models are also within error range.

\section{Uncovering hidden repeating FRBs}
\label{sec:uncover_repeaters}

\subsection{Combining models}
\label{subsec:combine_models}

After training the models, we apply the models back to the entire CHIME catalog to find the bursts reported as non-repeating in the catalog but are predicted as repeating based on our machine learning models. Because the training-test set split introduces randomness to the results, we repeat the splitting and training process 1000 times. In each trial, if one apparently non-repeating bursts is predicted as repeating by at least 4 of the 6 models described in Section \ref{sec:results} except the simple decision tree method, we record this occurrence. We further require that one burst needs to be predicted as a repeater in at least 100 trials to be identified as a repeater candidate.

The repeater candidates identified by this study, along with their input features and the number of trials predicted as repeating, are shown in Table \ref{table:candidates}.

\begin{table*}
    \hspace*{-0.45cm}
    \addtolength{\tabcolsep}{-2pt}
	\begin{tabular}{cccccccccccccccc}
		\hline
		\multirow{2}{*}{Name} & Sub & RA & Dec & $\Delta t_{BC}$ & $S_{\nu}$ & $F_{\nu}$ & $DM_E$ & $z$ & $\nu_c$ & $\log T_B$ & $\Delta t_r$ & $\log \Delta\nu$ & $\log E$ & \multirow{2}{*}{Score} & \multirow{2}{*}{Match}\\
		& Num & $^\circ$ & $^\circ$ & $\si{\milli\s}$ & $\si{\jansky}$ & $\si{\jansky\milli\s}$ & $\si{\parsec\per\cubic\centi\meter}$ & & $\si{\mega\hertz}$ & $\log\si{\kelvin}$ & $\si{\milli\s}$ & $\log\si{\mega\hertz}$ & $\log\si{\erg}$ & &\\
		\hline
        FRB20181229B & 0 & 238.37 & 19.78 & 20.64 & 0.42 & 4.9 & 359.7 & 0.32 & 445.5 & 33.093 & 2.546 & 2.19 & 39.773 & 320 &  \cmark\\
        \fbox{FRB20190423B} & 0 & 298.58 & 26.19 & 9.83 & 0.87 & 7.0 & 102.3 & 0.003 & 537.6 & 29.813 & 2.482 & 2.204 & 35.931 & 316 &  \cmark+\\
        FRB20190410A & 0 & 263.47 & -2.37 & 6.88 & 1.59 & 5.8 & 155.5 & 0.073 & 515.7 & 33.179 & 0.941 & 2.262 & 38.59 & 307 &  \cmark\\
        \fbox{FRB20181017B} & 0 & 237.76 & 78.5 & 12.78 & 1.06 & 6.5 & 263.7 & 0.207 & 593.2 & 33.271 & 1.914 & 2.394 & 39.628 & 301 &  \cmark+\\
        FRB20181128C & 0 & 268.77 & 49.71 & 14.75 & 0.39 & 3.4 & 569.2 & 0.557 & 480.3 & 33.785 & 1.477 & 2.42 & 40.144 & 301 &  \\
        FRB20190422A & 1 & 48.56 & 35.15 & 29.49 & 0.6 & 9.1 & 372.8 & 0.335 & 612.3 & 32.703 & 1.73 & 2.495 & 40.221 & 299 &  \cmark\\
        FRB20190409B & 0 & 126.65 & 63.47 & 30.47 & 0.39 & 6.8 & 237.8 & 0.175 & 545.5 & 32.008 & 1.991 & 2.528 & 39.464 & 296 &  \cmark\\
        \fbox{FRB20190329A} & 0 & 65.54 & 73.63 & 11.8 & 0.52 & 2.24 & 100.8 & 0.002 & 432.3 & 29.344 & 1.038 & 1.868 & 35.066 & 295 &  \cmark+\\
        \fbox{FRB20190423B} & 1 & 298.58 & 26.19 & 9.83 & 0.87 & 7.0 & 102.3 & 0.003 & 524.6 & 29.834 & 8.474 & 2.173 & 35.921 & 284 &  \cmark+\\
        \fbox{FRB20190206A} & 0 & 244.85 & 9.36 & 5.9 & 1.4 & 9.1 & 146.9 & 0.062 & 534.5 & 33.081 & 0.757 & 2.33 & 38.656 & 278 &  \cmark+\\
        FRB20190128C & 0 & 69.8 & 78.94 & 15.73 & 0.71 & 5.9 & 239.3 & 0.177 & 491.6 & 32.942 & 5.233 & 2.377 & 39.366 & 276 &  +\\
        FRB20190106A & 0 & 22.19 & 46.12 & 6.88 & 0.27 & 0.81 & 251.2 & 0.192 & 800.2 & 32.887 & 0.789 & 2.563 & 38.786 & 270 &  \\
        FRB20190129A & 0 & 45.06 & 21.42 & 8.85 & 0.49 & 5.0 & 432.7 & 0.404 & 707.7 & 33.703 & 0.805 & 2.541 & 40.191 & 270 &  \cmark\\
        FRB20181030E & 0 & 135.67 & 8.89 & 5.9 & 2.0 & 6.3 & 109.8 & 0.013 & 470.5 & 31.992 & 0.395 & 2.222 & 37.086 & 264 &  \cmark\\
        \fbox{FRB20190527A} & 0 & 12.45 & 7.99 & 57.02 & 0.47 & 10.1 & 550.9 & 0.537 & 484.7 & 32.651 & 1.737 & 2.313 & 40.588 & 261 &  \cmark+\\
        FRB20190218B & 0 & 268.7 & 17.93 & 17.69 & 0.57 & 5.9 & 466.3 & 0.442 & 588.0 & 33.409 & 1.422 & 2.524 & 40.264 & 259 &  \cmark\\
        FRB20190609A & 1 & 345.3 & 87.94 & 4.92 & 3.6 & 10.4 & 258.2 & 0.2 & 600.5 & 34.591 & 1.767 & 2.428 & 39.808 & 244 &  \cmark\\
        \fbox{FRB20190412B} & 0 & 285.65 & 19.25 & 42.27 & 0.68 & 12.8 & 110.9 & 0.015 & 400.2 & 30.045 & 6.702 & 2.359 & 37.416 & 239 &  \cmark+\\
        FRB20190125B & 0 & 231.45 & 50.54 & 6.88 & 0.83 & 4.7 & 145.0 & 0.059 & 609.9 & 32.57 & 2.332 & 2.627 & 38.391 & 239 &  +\\
        \fbox{FRB20181231B} & 0 & 128.77 & 55.99 & 2.95 & 0.89 & 2.34 & 150.3 & 0.066 & 657.7 & 33.366 & 0.316 & 2.442 & 38.217 & 229 &  \cmark+\\
        FRB20181221A & 0 & 230.58 & 25.86 & 4.92 & 1.25 & 5.8 & 291.8 & 0.24 & 510.1 & 34.438 & 0.608 & 2.231 & 39.648 & 187 &  \cmark\\
        FRB20190112A & 0 & 257.98 & 61.2 & 9.83 & 1.4 & 16.2 & 383.8 & 0.348 & 697.7 & 33.946 & 1.217 & 2.502 & 40.562 & 176 &  \cmark\\
        FRB20190125A & 0 & 45.73 & 27.81 & 13.76 & 0.37 & 2.6 & 504.3 & 0.484 & 655.5 & 33.428 & 2.162 & 2.634 & 40.038 & 174 &  \\
        FRB20181218C & 0 & 285.98 & 58.24 & 4.92 & 0.25 & 1.73 & 319.8 & 0.273 & 472.7 & 33.92 & 4.084 & 2.608 & 39.205 & 147 &  \\
        \fbox{FRB20190429B} & 0 & 329.93 & 3.96 & 16.71 & 0.74 & 5.0 & 253.5 & 0.194 & 422.4 & 33.122 & 5.342 & 1.705 & 39.311 & 138 &  \cmark+\\
        \fbox{FRB20190109B} & 0 & 253.47 & 1.25 & 6.88 & 1.2 & 3.0 & 106.9 & 0.009 & 408.1 & 31.455 & 0.337 & 1.969 & 36.397 & 137 &  \cmark+\\
        FRB20190206B & 0 & 49.76 & 79.5 & 19.66 & 0.95 & 9.6 & 274.0 & 0.219 & 506.4 & 33.039 & 5.824 & 2.544 & 39.78 & 126 &  \\
		\hline
	\end{tabular}
	\addtolength{\tabcolsep}{2pt}
	\caption{List of repeater candidates from the combined models. Check mark \cmark\;indicates this FRB is also identified as a repeater candidate by \citet{zhu-ge2022MachineLearningClassification}, plus mark + indicates this FRB is also identified as a repeater candidate by the leave-one-out $k$-NN method. The strongest candidates are marked with boxes around their names.}
	\label{table:candidates}
\end{table*}

\subsection{Leave-one-out \textit{k}-nearest neighbor}
\label{subsec:leave_one_out KNN}

Another way to identify repeater candidates is Leave-one-out $k$-nearest neighbor. $k$-nearest neighbour ($k$-NN) classification \citep{fix1989DiscriminatoryAnalysisNonparametric,altman1992IntroductionKernelNearestNeighbor} takes the training set and project the feature vectors into a hyperspace. Then, for any prediction input, the model finds the nearest $k$ training data points to the input feature vector. The prediction output label is then determined by the majority label of the nearest neighbors. In this study, we use $k=5$.

In each cycle, one of the non-repeating burst is excluded from the training set, which consists of all the rest of the data. Then, the one left out non-repeating burst is predicted with the trained model. If the output label is repeating, then this burst is identified as a repeater candidate. This method essentially looks at the ``surrounding" bursts in the feature hyperparameter to determine if the left-out non-repeater is closer to the repeaters. The results are shown in Table \ref{table:candidates_knn}.

\subsection{Unsupervised learning and the strongest candidates}
\label{subsec:unsupervised}

The classification results using unsupervised learning are reported in a companion paper \citep{zhu-ge2022MachineLearningClassification}. Those methods also identified a list of repeater candidates which overlap with our predicted repeater candidates. In Table \ref{table:candidates}, we have highlighted the strongest candidates that are identified by all three methods, namely the combined supervised model, the leave-one-out KNN method and the unsupervised machine learning methods (enclosed in boxes). These candidates can be regarded as strong candidate targets for searching for repeated bursts.

\begin{table*}
    \addtolength{\tabcolsep}{-2pt}
    \begin{tabular}{cccccccccccccc}
		\hline
		\multirow{2}{*}{Name} & Sub & RA & Dec & $\Delta t_{BC}$ & $S_{\nu}$ & $F_{\nu}$ & $DM_E$ & $z$ & $\nu_c$ & $\log T_B$ & $\Delta t_r$ & $\log \Delta\nu$ & $\log E$\\
		& Num & $^\circ$ & $^\circ$ & $\si{\milli\s}$ & $\si{\jansky}$ & $\si{\jansky\milli\s}$ & $\si{\parsec\per\cubic\centi\meter}$ & & $\si{\mega\hertz}$ & $\log\si{\kelvin}$ & $\si{\milli\s}$ & $\log\si{\mega\hertz}$ & $\log\si{\erg}$ \\
		\hline
        FRB20180801A & 0 & 322.53 & 72.72 & 9.83 & 1.11 & 7.9 & 565.6 & 0.553 & 595.6 & 34.399 & 0.373 & 2.32 & 40.597 \\
        FRB20180925A & 0 & 74.93 & 77.99 & 14.75 & 0.99 & 8.7 & 167.1 & 0.088 & 800.2 & 32.093 & 4.156 & 2.602 & 39.12 \\
        FRB20180928A & 0 & 312.95 & 30.85 & 2.95 & 1.34 & 2.5 & 94.7 & 0.002 & 400.2 & 31.027 & 0.268 & 1.963 & 35.08 \\
        FRB20181017B & 0 & 237.76 & 78.5 & 12.78 & 1.06 & 6.5 & 263.7 & 0.207 & 593.2 & 33.271 & 1.914 & 2.313 & 39.628 \\
        FRB20181119B & 0 & 299.38 & 31.12 & 22.61 & 4.5 & 94.0 & 166.8 & 0.087 & 536.4 & 32.723 & 3.063 & 2.602 & 39.976 \\
        FRB20181231B & 0 & 128.77 & 55.99 & 2.95 & 0.89 & 2.34 & 150.3 & 0.066 & 657.7 & 33.366 & 0.316 & 2.414 & 38.217 \\
        FRB20190107A & 0 & 0.86 & 21.81 & 47.19 & 0.49 & 6.3 & 809.3 & 0.824 & 400.2 & 33.377 & 14.198 & 2.426 & 40.678 \\
        FRB20190109B & 0 & 253.47 & 1.25 & 6.88 & 1.2 & 3.0 & 106.9 & 0.009 & 408.1 & 31.455 & 0.337 & 1.965 & 36.397 \\
        FRB20190124E & 0 & 297.75 & 20.57 & 18.68 & 0.64 & 7.3 & 225.8 & 0.161 & 625.6 & 32.451 & 4.945 & 2.602 & 39.476 \\
        FRB20190125B & 0 & 231.45 & 50.54 & 6.88 & 0.83 & 4.7 & 145.0 & 0.059 & 609.9 & 32.57 & 2.332 & 2.602 & 38.391 \\
        FRB20190128C & 0 & 69.8 & 78.94 & 15.73 & 0.71 & 5.9 & 239.3 & 0.177 & 491.6 & 32.942 & 5.233 & 2.307 & 39.366 \\
        FRB20190206A & 0 & 244.85 & 9.36 & 5.9 & 1.4 & 9.1 & 146.9 & 0.062 & 534.5 & 33.081 & 0.757 & 2.304 & 38.656 \\
        FRB20190329A & 0 & 65.54 & 73.63 & 11.8 & 0.52 & 2.24 & 100.8 & 0.002 & 432.3 & 29.344 & 1.038 & 1.867 & 35.066 \\
        FRB20190412B & 0 & 285.65 & 19.25 & 42.27 & 0.68 & 12.8 & 110.9 & 0.015 & 400.2 & 30.045 & 6.702 & 2.353 & 37.416 \\
        FRB20190423B & 0 & 298.58 & 26.19 & 9.83 & 0.87 & 7.0 & 102.3 & 0.003 & 537.6 & 29.813 & 2.482 & 2.202 & 35.931 \\
        FRB20190423B & 1 & 298.58 & 26.19 & 9.83 & 0.87 & 7.0 & 102.3 & 0.003 & 524.6 & 29.834 & 8.474 & 2.172 & 35.921 \\
        FRB20190429B & 0 & 329.93 & 3.96 & 16.71 & 0.74 & 5.0 & 253.5 & 0.194 & 422.4 & 33.122 & 5.342 & 1.627 & 39.311 \\
        FRB20190527A & 0 & 12.45 & 7.99 & 57.02 & 0.47 & 10.1 & 550.9 & 0.537 & 484.7 & 32.651 & 1.737 & 2.126 & 40.588 \\
        FRB20190617B & 0 & 56.43 & 1.16 & 13.76 & 0.99 & 9.2 & 229.7 & 0.166 & 459.3 & 33.201 & 6.504 & 2.271 & 39.469 \\
		\hline
	\end{tabular}
	\addtolength{\tabcolsep}{2pt}
	\caption{List of repeater candidates from leave-one-out KNN.}
	\label{table:candidates_knn}
\end{table*}

\section{Conclusions}
\label{sec:conclusions}

In this paper, we have applied an array of supervised machine learning methods to the first CHIME FRB catalog, with the aim of categorizing repeating and non-repeating FRBs and identifying repeater candidates from the apparently non-repeating sample. We reach the following conclusions:
\begin{itemize}
    \item The parameter distributions of repeating and non-repeating FRBs are significantly different from each other. Machine learning algorithms, even some simple ones, are able to distinguish them. The machine learning algorithms confirm the finding of the CHIME team \citep{chime/frbcollaboration2021FirstCHIMEFRB} that spectral bandwidth is one key parameter to distinguish between the two categories. Furthermore, several algorithms identify brightness temperature (with a proper $(1+z)$ correction, see Eq.(\ref{eq:TB}) and Appendix \ref{sec:derivation_T_B}) as another key parameter to distinguish the two categories. 
    \item The SVM model performs the best in predicting repeating and non-repeating FRBs, while other models are not significantly far behind.
    \item The fact that the machine learning algorithms can correctly predict the majority of the bursts hints that there likely exist two distinct types of FRBs. Most of the apparently non-repeating FRBs may either have a distinct type of central engine or a distinct radiation mechanism that is different from that of repeating FRBs. 
    \item Some non-repeating FRBs are predicted as repeating FRBs by multiple models, suggesting they might be hidden repeaters yet to be uncovered. Many candidates are also identified from unsupervised machine learning methods reported in Paper II. We have identified several strong candidates by cross comparing the candidates from the two papers (Table \ref{table:candidates}) and recommend the community to search for repeated bursts from these sources. The algorithms applied in this paper can also be applied to newly discovered FRBs to predict whether they will repeat, which will help to arrange follow-up observations accordingly.
\end{itemize}

\section{Acknowledgements}
\label{sec:acknowledgements}

The authors thank the anonymous referee for comments, Dongzi Li and Kiyoshi Masui for the information on the light curve and spectral parameters presented in the first CHIME FRB catalog, and the UNLV transient group members Shunke Ai, Connery Chen, Emily Huerta, and Yuanhong Qu, as well as Weiyang Wang for various discussions, especially on the derivations of FRB brightness temperature. Dian Shi, Paul La Plante and Beiyu Lin are acknowledged for giving helpful advice on data pre-processing and machine learning model constructions. This work is partially supported by the Top Tier Doctoral Graduate Research Assistantship (TTDGRA) and Nevada Center for Astrophysics at the University of Nevada, Las Vegas. 

\section*{Data availability}
\label{sec:data_availability}
The CHIME/FRB catalog used in this paper is public and is available at \url{https://www.chime-frb.ca/catalog}. The code used in this study is available as supplementary material in the online version.

\bibliographystyle{mnras}
\bibliography{frb-ml}

\appendix

\section{Derivation of brightness temperature \texorpdfstring{$T_B$}{T\_B}}
\label{sec:derivation_T_B}

In the literature, the derivation of $T_B$ for FRBs has not properly considered the redshift correction \citep{zhang2020PhysicalMechanismsFast, xiao2021PhysicsFastRadio, xiao2022NewInsightsCriterion}. We suggest that the proper correction is Equation (\ref{eq:TB}). In the following, we present two independent proofs to justify it.

\subsection{The \texorpdfstring{$D_{\rm A}$}{D\_A} approach}

For an FRB source with specific intensity $I_{\nu_0}$ at the cosmic rest-frame frequency $\nu_0$, the brightness temperature $T_B$ at $\nu_0$ can be defined using the Rayleigh-Jeans approximation of the blackbody law:
\begin{equation}
    I_{\nu_0}\dd\nu_0=\frac{2\nu_0 k_B T_B}{c^2}\dd\nu_0,
    \label{eq:Inu0}
\end{equation}
where $k_B$ is Boltzmann constant.

Suppose the observed specific intensity is $I_\nu$ and the observed frequency is $\nu = \nu_0/(1+z)$. Since $I_\nu / \nu^3$ should remain constant \citep{rybicki1991RadiativeProcessesAstrophysics}, one can write
\begin{equation}
    I_{\nu}\dd\nu=\frac{I_{\nu_0}\dd\nu_0}{(1+z)^4}=\frac{2\nu_0 k_B T_B}{c^2(1+z)^3}\dd\nu.
    \label{eq:Inu}
\end{equation}

The observed solid angle of the FRB emission region can be written as
\begin{equation}
    \Delta\Omega = \frac{\upi R^2}{D_{\rm A}^2}
= \frac{\upi (c\Delta t_0)^2}{D_{\rm A}^2},
\end{equation}
where $R$ is the radius of the emission region estimated by the light travel distance in the duration of the FRB, $R=c\Delta t_0$, $\Delta t_0$ is the duration of the FRB at the source, and $D_A$ is the angular diameter distance of the FRB. 

Finally, the observed specific flux $S_{\nu}$ can be expressed as
\begin{equation}
    S_{\nu}\dd\nu = \Delta\Omega I_{\nu}\dd\nu
    \label{eq:Snu}
\end{equation}
Substituting Eq.(\ref{eq:Inu}) to Eq.(\ref{eq:Snu}), one finally obtains
\begin{align}
    T_B&=\frac{S_\nu D_{\rm A}^2} {2\upi k_B (\nu \Delta t)^2} (1+z)^3 \nonumber\\
    &=\SI{1.1e35}{\kelvin} \qty(\frac{S_{\nu}}{\si{\jansky}}) \qty(\frac{\nu}{\si{\giga\hertz}})^{-2} \qty(\frac{\Delta t}{\si{\milli\s}})^{-2} \qty(\frac{D_{\rm A}}{\si{\giga\parsec}})^{2}(1+z)^3
\end{align}
Where $\Delta t=(1+z)\Delta t_0$ is observed duration, and the $(1+z)$ factors from $\Delta t$ and $\nu$ cancel out. This is one expression in Eq.(\ref{eq:TB}).

\subsection{The \texorpdfstring{$D_{\rm L}$}{D\_L} approach}
Another route to derive this expression of brightness temperature is similar to the derivation of the Stefan-Boltzmann law. The specific luminosity of the FRB can be expressed as
\begin{equation}
    L_{\nu_0}\dd\nu_0=4\upi R^2 \upi I_{\nu_0}\dd\nu_0.
\end{equation}

From the definition of luminosity distance $D_{\rm L}$, the observed specific flux $S_{\nu}$ can be written as:
\begin{equation}
    S_{\nu}\dd\nu=\frac{L_{\nu_0}\dd\nu_0}{4\upi D_{\rm L}^2}
    \label{eq:Snu2}
\end{equation}
Substituting Eq.(\ref{eq:Inu0}) to Eq.(\ref{eq:Snu2}) and noting  $\dd\nu_0=(1+z)\dd\nu$, one finally obtains
\begin{align}
    T_B&=\frac{S_\nu D_{\rm L}^2} {2\upi k_B (\nu \Delta t)^2}\frac{1}{1+z} \nonumber\\
    &=\SI{1.1e35}{\kelvin} \qty(\frac{S_{\nu}}{\si{\jansky}}) \qty(\frac{\nu}{\si{\giga\hertz}})^{-2} \qty(\frac{\Delta t}{\si{\milli\s}})^{-2} \qty(\frac{D_{\rm L}}{\si{\giga\parsec}})^{2}\frac{1}{1+z}.
\end{align}
This is another expression in Eq.(\ref{eq:TB}). Since the two distances are connected through $D_{\rm L}=D_{\rm A}(1+z)^2$ for a flat universe, the two expressions are equivalent and the two derivations are consistent with each other.

\section{List of parameters used in machine learning models}
\label{sec:list_of_parameter}

\bottomcaption{List of parameters used in the machine learning models.}
\label{table:paramters_list}
\begin{supertabular}{ll}
		\hline
		Name & Value \\
		\hline
		\multicolumn{2}{c}{Decision Tree} \\
		\hline
        \texttt{criterion} & `gini' \\
        \texttt{splitter} & `best' \\
        \texttt{max\_depth} (all features) & 5 \\
        \texttt{max\_depth} ($T_B$ and $\Delta\nu$) & 3 \\
        \texttt{min\_samples\_split} & 2 \\
        \texttt{min\_samples\_leaf} & 1 \\
        \texttt{min\_weight\_fraction\_leaf} & 0.0 \\
        \texttt{max\_features} & None \\
        \texttt{max\_leaf\_nodes} & None \\
        \texttt{min\_impurity\_decrease} & 0.0 \\
        \texttt{class\_weight} & None \\
        \texttt{ccp\_alpha} & 0.0 \\
        \hline
		\multicolumn{2}{c}{Random Forest} \\
		\hline
		\texttt{n\_estimators} & 100 \\
        \texttt{criterion} & `gini' \\
        \texttt{max\_depth} & None \\
        \texttt{min\_samples\_split} & 2 \\
        \texttt{min\_samples\_leaf} & 1 \\
        \texttt{min\_weight\_fraction\_leaf} & 0.0 \\
        \texttt{max\_features} & `sqrt' \\
        \texttt{max\_leaf\_nodes} & None \\
        \texttt{min\_impurity\_decrease} & 0.0 \\
        \texttt{bootstrap} & True \\
        \texttt{oob\_score} & False \\
        \texttt{class\_weight} & None \\
        \texttt{ccp\_alpha} & 0.0 \\
        \texttt{max\_samples} & None \\
        \hline
		\multicolumn{2}{c}{AdaBoost} \\
		\hline
		\texttt{base\_estimator} & None \\
        \texttt{n\_estimators} & 50 \\
        \texttt{learning\_rate} & 1.0 \\
        \texttt{algorithm} & `SAMME.R' \\
        \hline
		\multicolumn{2}{c}{LightGBM} \\
		\hline
		\texttt{boosting\_type} & `gbdt' \\
        \texttt{num\_leaves} & 31 \\
        \texttt{max\_depth} & -1 \\
        \texttt{learning\_rate} & 0.1 \\
        \texttt{n\_estimators} & 100 \\
        \texttt{subsample\_for\_bin} & 200000 \\
        \texttt{objective} & None \\
        \texttt{class\_weight} & None \\
        \texttt{min\_split\_gain} & 0 \\
        \texttt{min\_child\_weight} & 1e-3 \\
        \texttt{min\_child\_samples} & 20 \\
        \texttt{subsample} & 1 \\
        \texttt{subsample\_freq} & 0 \\
        \texttt{colsample\_bytree} & 1 \\
        \texttt{reg\_alpha} & 0 \\
        \texttt{reg\_lambda} & 0 \\
        \texttt{importance\_type} & `split' \\
		\hline
		\multicolumn{2}{c}{XGBoost} \\
		\hline
		\texttt{objective} & `binary:logistic' \\
        \texttt{use\_label\_encoder} & False \\
        \texttt{n\_estimators} & 100 \\
        \texttt{max\_depth} & 6 \\
        \texttt{max\_leaves} & 0 \\
        \texttt{max\_bin} & 256 \\
        \texttt{grow\_policy} & `depthwise' \\
        \texttt{learning\_rate} & 0.3 \\
        \texttt{booster} & `gbtree' \\
        \texttt{tree\_method} & `exact' \\
        \texttt{gamma} & 0 \\
        \texttt{min\_child\_weight} & 1 \\
        \texttt{max\_delta\_step} & 0 \\
        \texttt{subsample} & 1 \\
        \texttt{sampling\_method} & `uniform' \\
        \texttt{colsample\_bytree} & 1 \\
        \texttt{colsample\_bylevel} & 1 \\
        \texttt{colsample\_bynode} & 1 \\
        \texttt{reg\_alpha} & 0 \\
        \texttt{reg\_lambda} & 1 \\
        \texttt{scale\_pos\_weight} & 1 \\
        \texttt{base\_score} & 0.5 \\
        \texttt{num\_parallel\_tree} & 1 \\
        \texttt{monotone\_constraints} & `()' \\
        \texttt{interaction\_constraints} & `' \\
        \texttt{importance\_type} & None \\
        \texttt{predictor} & `auto' \\
        \texttt{enable\_categorical} & False \\
        \texttt{max\_cat\_to\_onehot} & 4 \\
        \texttt{eval\_metric} & None \\
        \texttt{early\_stopping\_rounds} & None \\
        \texttt{callbacks} & None \\
		\hline
		\multicolumn{2}{c}{SVM} \\
		\hline
		\texttt{C} & 1.0 \\
        \texttt{kernel} & `rbf' \\
        \texttt{degree} & 3 \\
        \texttt{gamma} & `scale' \\
        \texttt{coef0} & 0.0 \\
        \texttt{shrinking} & True \\
        \texttt{probability} & False \\
        \texttt{tol} & 0.001 \\
        \texttt{class\_weight} & None \\
        \texttt{max\_iter} & -1 \\
        \texttt{decision\_function\_shape} & `ovr' \\
        \texttt{break\_ties} & False \\
		\hline
		\multicolumn{2}{c}{Nearest Centroid} \\
		\hline
		\texttt{metric} & `euclidean' \\
		\texttt{shrink\_threshold} & None \\
		\hline
\end{supertabular}
\bsp
\label{lastpage}
\end{document}